\documentclass[12pt]{article}
\usepackage{epsf} 

\begin{document}

\newcommand{\sheptitle}
{String Thermodynamics in D-Brane Backgrounds}

\newcommand{\shepauthor}
{
S.A.~Abel$^{a,}$\footnote[1]{Steven.Abel@cern.ch}, 
J.L.F.~Barb\'on$^{b,}$\footnote[2]{J.Barbon@phys.uu.nl}, 
I.I.~Kogan$^{c,}$\footnote[3]{i.kogan@physics.ox.ac.uk}, 
E.~Rabinovici$^{d,}$\footnote[4]{ELIEZER@vms.huji.ac.il}}

\newcommand{\shepaddress}
{
$a$ Theory Division, CERN, CH-1211 Geneva 23, Switzerland\\
$b$ Spinoza Institute, Leuvenlaan 4, 3584 CE, Utrecht, The Netherlands\\
$c$ Theoretical Physics, 1 Keble Road, Oxford OX1 3NP, UK\\
$d$ Racah Institute of Physics, The Hebrew University, Jerusalem, Israel}

\newcommand{\shepabstract}
{We discuss the thermal properties of string gases
propagating in various D-brane backgrounds in the weak-coupling limit,
and at temperatures close to the Hagedorn temperature. 
We determine, in the canonical ensemble, whether the Hagedorn 
temperature is limiting or non-limiting. This depends on the 
dimensionality of 
the D-brane, and the size of the compact dimensions.
We find that in many cases the non-limiting behaviour manifest
in the canonical ensemble is modified to a limiting behaviour
in the 
microcanonical ensemble and show that, 
when there are different systems in thermal contact, the energy 
flows into open strings on the `limiting' D-branes of largest 
dimensionality. Such energy densities may eventually exceed the
D-brane intrinsic tension. We discuss possible implications of this
for the survival of D$p$-branes with large values of $p$ in an
early cosmological Hagedorn regime. We also discuss the general 
phase diagram of the interacting theory, as implied by the
holographic and black-hole/string correspondence principles.                  
}

\begin{titlepage}
\begin{flushright}
CERN-TH/98-375\\
OUTP-98-81-P\\
SPIN-98-7\\
hep-th/9902058\\

\end{flushright}
\vspace{0.5in}
\begin{center}
{\large{\bf \sheptitle}}
\bigskip \\ \shepauthor \\ \mbox{} \\ {\it \shepaddress} \\ 
\vspace{0.5in}
{\bf Abstract} \bigskip \end{center} \setcounter{page}{0}
\shepabstract
\vspace{0.5in}
\begin{flushleft}
CERN-TH/98-375\\
\today
\end{flushleft}
\end{titlepage}

\def\sspace{\baselineskip = .16in}
\def\dspace{\baselineskip = .30in}
\def\beq{\begin{equation}}
\def\eeq{\end{equation}}
\def\bea{\begin{eqnarray}}
\def\eea{\end{eqnarray}}
\def\bq{\begin{quote}}
\def\eq{\end{quote}}
\def\ra{\rightarrow}
\def\lra{\leftrightarrow}
\def\ups{\upsilon}
\def\bq{\begin{quote}}
\def\eq{\end{quote}}
\def\ra{\rightarrow}
\def\un{\underline}
\def\ov{\overline}
\def\ord{{\cal O}} 

\newcommand{\plb}[3]{{{\it Phys.~Lett.}~{\bf B#1} (#3) #2}}
\newcommand{\npb}[3]{{{\it Nucl.~Phys.}~{\bf B#1} (#3) #2}}
\newcommand{\prd}[3]{{{\it Phys.~Rev.}~{\bf D#1} (#3) #2}}
\newcommand{\ptp}[3]{{{\it Prog.~Theor.~Phys.}~{\bf #1} (#3) #2}}
\newcommand{\ijmpa}[3]{{{\it Int.~J.~Mod.~Phys.}~{\bf A#1} (#3) #2}}
\newcommand{\prl}[3]{{{\it Phys.~Rev.~Lett.}~{\bf #1} (#3) #2}}
\newcommand{\hepph}[1]{{\tt hep-ph/#1}}
\newcommand{\hepth}[1]{{\tt hep-th/#1}}
\newcommand{\grqc}[1]{{\tt gr-qc/#1}} 
\newcommand{\leqsim}{\,\raisebox{-0.6ex}{$\buildrel < \over \sim$}\,}
\newcommand{\geqsim}{\,\raisebox{-0.6ex}{$\buildrel > \over \sim$}\,}
\newcommand{\be}{\begin{equation}}
\newcommand{\ee}{\end{equation}}
\newcommand{\ba}{\begin{eqnarray}}
\newcommand{\ea}{\end{eqnarray}}
\newcommand{\nn}{\nonumber}
\newcommand{\cf}{\mbox{{\em c.f.~}}}
\newcommand{\ie}{\mbox{{\em i.e.~}}}
\newcommand{\eg}{\mbox{{\em e.g.~}}}
\newcommand{\mpl}{\mbox{$M_{pl}$}}
\newcommand{\ol}[1]{\overline{#1}}
\def\gev{\,{\rm GeV }}
\def\tev{\,{\rm TeV }}
\def\dd{\mbox{d}}
\def\etal{\mbox{\it et al }}
\def\half{\frac{1}{2}}
\def\Tr{\mbox{Tr}}
\def\bra{\langle}
\def\ket{\rangle}
\def\lim{\mbox{{\bf L}} }
\def\nlim{\mbox{{\bf NL}} }
\def\sclim{\mbox{\tiny{\bf L}} }
\def\scnlim{\mbox{\tiny{\bf NL}} }
\def\nlimc{\mbox{{\bf NL$_{closed}$}} }
\def\nlimo{\mbox{{\bf NL$_{open}$}} }
\def\Vp{V_{\parallel}}
\def\Vt{V_{\perp}} 
\def\Rt{R_{\perp}}
\def\ep{\varepsilon} 
\newcommand{\smallfrac}[2]{\frac{\mbox{\small #1}}{\mbox{\small #2}}}

\def\CAG{{\cal A/\cal G}}           \def\CO{{\cal O}} \def\CZ{{\cal Z}}
\def\CA{{\cal A}} \def\CC{{\cal C}} \def\CF{{\cal F}} \def\CG{{\cal G}}
\def\CL{{\cal L}} \def\CH{{\cal H}} \def\CI{{\cal I}} \def\CU{{\cal U}}
\def\CB{{\cal B}} \def\CR{{\cal R}} \def\CD{{\cal D}} \def\CT{{\cal T}}
\def\CM{{\cal M}} \def\CP{{\cal P}}
\def\CN{{\cal N}} \def\CS{{\cal S}}

\newpage 

\section{Introduction and Background}

Models in which the particle spectrum has the Hagedorn form~\cite{hag},  
\be 
\label{eq:hag}
\rho(m) \sim  m^a \,\,e^{bm}
\ee
are of great interest because of their  thermodynamic properties. 
For example, for $a<-5/2$ (in four dimensions) and, at sufficient 
energy density, a system like this has a 
negative specific heat. Thermodynamic quantities 
are not extensive and two such sytems cannot establish an
equilibrium~\cite{carlitz}. 

This type of spectrum first arose in the context of statistical
bootstrap models~\cite{hag,duals} and, for
 hadrons, such behaviour indicates that they are composed of more 
fundamental constituents \cite{cab}. In  
fundamental string theories we find  the same kind of 
spectrum~\cite{others,bowick,brandvafa,deo}, and a search for hints to the
existence of `string constituents' is of great interest. 

On a more practical level, regimes of Hagedorn behaviour 
of weakly-coupled strings are interesting in the context
of stringy cosmological models \cite{brandvafa}.
 In particular, there has been much work  recently in  
models where the string scale can be significantly lower than the
Planck scale~\cite{hw}, perhaps even as low as the 
\tev scale~\cite{tev,dienes}.\footnote{Models
 with closed-string winding modes
at the \tev scale were proposed in \cite{anto}.} 
 These models
involve string theory in backgrounds in which
the gauge sector is  confined to 
extended  topological defects (branes) of various kinds.  
Particularly tractable are models constructed 
with Dirichlet $p$-branes~\cite{polch}. 
  The visible universe could, for example, 
correspond to a D3-brane,  and the
 cosmological behaviour of such
  systems is only  beginning to be  studied ~\cite{tev,dienes,tevcosmo}.

In this paper we study various aspects of the thermodynamics of 
fundamental strings in  backgrounds with webs of intersecting
 D-branes. For any particular brane structure (\ie number 
and spatial arrangement of the D-branes), there are
open strings in different  sectors, labeled by the  D-brane sets
 to which they are attached, as well as
closed strings propagating in the bulk.  
We determine the thermodynamic properties of the different D-brane sectors
at energy densities larger than the fundamental string scale,
  to leading order in string
perturbation theory, and paying special attention to the dependence on
the various T-moduli (volumes).
For early work on various aspects of
Hagedorn behaviour with D-branes see for example
\cite{greent,thor,mav,jose,us}.
 
One particularly interesting fact is that for 
ordinary ten-dimensional superstrings
(including the heterotic) the closed-string sector has a Hagedorn
temperature which is `non-limiting' (in that it requires a 
finite amount of energy 
to reach it, in the description provided by the canonical ensemble),
 whilst Type--I open strings  have a
 `limiting' Hagedorn temperature~\cite{others,dienes}. It was pointed out
recently in Ref. \cite{us} that open-string sectors in D$p$-branes show  
`limiting' behaviour provided $p\geq 5$. On the other hand, for $p<5$, open
strings seem to show `non-limiting' behaviour, similar to that of closed
strings.  It should be noted    
that    different
sectors have the same Hagedorn temperature in perturbation theory,  
since the critical behaviour  can be 
 related to  the onset of infrared divergences due to a 
{\it closed}-string state becoming massless at the Hagedorn temperature
 \cite{ian,awit}.
Provided this `tachyonic' 
closed-string state couples to all D-branes, all the  topologically
distinct open-string sectors will share the same critical temperature.  

We begin our discussion in section 2 by calculating
the canonical (single-string) density of states of an open string
propagating in various D-brane backgrounds. In particular we 
generalize the analysis 
to the case where the dimensions are large 
but compact and pay special attention to whether
  the Hagedorn temperature 
appears to be `limiting' or `non-limiting'. 
When dealing with finite and large dimensions, 
the experience with closed strings
(\cf \cite{deo}) 
tells us that the thermodynamic properties ought to change 
as the energy is raised through `thresholds'. These thresholds
 correspond to the string being able to `feel' extra dimensions
by producing winding or heavy momentum modes and we shall find 
that this is indeed the case with open strings.
Any dependence on the finite size of extra dimensions is of 
particular interest because phenomenologically viable
D-brane scenarios typically require large compact dimensions in order 
to explain why the weak scale is so much lower than the Planck scale.

In section 3 we  derive the thermodynamic 
properties in the microcanonical ensemble.
As with closed strings, this 
analysis is required once the canonical ensemble exhibits esoteric features
such as supposedly negative specific heat,
  and leads to a better understanding
of the thermodynamic properties. 
The more limited information encoded in the free energy (the canonical 
ensemble) concerning the properties of non-limiting strings is greatly 
enhanced by studying the microcanonical ensemble. 

Most importantly 
the universal presence of gravity in any string system means that the
infinite volume limit (the thermodynamic limit) at finite energy density
does not exist in a strict sense, due to the Jeans instability  
\cite{grossp,awit}, and the  holographic bound \cite{holo}. Thus, consistency
requires working in finite volume, and investigating whether
there are regimes of approximate thermodynamic behaviour for each 
individual case. 

To do this, we 
shall work in the simplest finite-volume backgrounds, \ie toroidal
compactifications.
 For closed strings in the ideal-gas approximation,
it was found in \cite{brandvafa,deo} that winding modes tend to work in favour of
positive specific heat. Indeed, if winding modes carry a sizeable proportion
of the energy, a superficially non-limiting behaviour according to the
canonical ensemble may turn into a limiting behaviour in the true
microcanonical analysis.

We find many examples of this phenomenon in the brane backgrounds. 
As the microcanonical discussion can be rather technical, 
it is worth previewing the resulting physical picture. 
Imagine heating up open-string excitations on a thermally isolated 
 D-brane wrapped on a finite-volume torus.   
Consider a D$p$-brane for which the canonical ensemble predicts 
a non-limiting
behaviour. 
Eventually the critical Hagedorn energy density is reached on the brane and 
open strings begin looping into the bulk volume although their ends 
must stay attached to the brane. The D-brane is now surrounded 
by an open-string cloud which spreads as we raise the temperature.
At some point a few energetic strings emerge and,
as we raise the temperature 
still further, the spectrum of a canonically non-limiting open-string
system becomes resolved into a peak of low-energy
excitations and a few energetic excitations which carry most of the 
energy. Eventually these modes are able to wind in the Dirichlet directions
and their number grows rapidly once they start winding. The thermal
properties begin to resemble those of the system in a small, 
totally compact volume. As we approach the Hagedorn temperature, the 
specific heat increases dramatically, and we find that we cannot supply 
enough energy to raise the temperature to the Hagedorn temperature.
The limiting behaviour has been restored.

In the more general multibrane configurations there are several types
of open strings depending on how these strings stretch between
branes attached at their end-points. We calculate the entropy for each
such class. We find that the critical behaviour is very similar in
all open-string sectors.  
 
The thermal interaction of two or more Hagedorn systems  
then follows directly (in section 4) from the microcanonical discussion and 
turns out to be quite unusual. In particular, we will show that 
systems which are `non-limiting' tend to give their energy up to 
`limiting' systems. Thus if we take our previously isolated D-brane 
and place it in a bath of closed strings, the energy of the former 
increases in the manner described above, almost without limit. 
This curious and possibly violent 
disequilibrium is due to differently diverging 
specific heats and is reminiscent of systems 
with negative specific heat 
(although we stress that most specific heats are found to be positive
below the Hagedorn temperature).

We then speculate on how such a process might end.
We suggest that eventually the energy density of the open-string gas 
becomes greater than the D-brane tension. At this point the system is
unstable towards the thermal nucleation of D-brane--antiD-brane pairs
of various dimensions and topological structures.
We make some estimates of the  production rate under the
assumption that the brane--antibrane pairs form a dilute plasma.  

Finally, in section 5, we 
suggest a  phase diagram  including the effects of string
interactions. Most notably, we use the correspondence principle of
\cite{polhor} to derive high-energy generalizations of previously
studied phase diagrams in the context of the SYM/AdS correspondence
\cite{malda,maldai,us,mart}. It is suggested that, at weak string coupling,
the Hagedorn regime is always bounded by a black-hole-dominated phase,
which subsequently saturates the holographic bound \cite{holo}.  
Indeed, black holes seem to emerge quite often when the Hagedorn
regime is probed \cite{bowick,ven,sus,polhor,us}. 

\section{The Canonical Ensemble in the 
Presence of D-branes} 

We shall consider models of Type--II strings on tori, with a number
of D-branes wrapped in a possibly complicated intersection pattern,
together with orientifold planes ensuring the appropriate cancellation
of tadpoles and anomalies.  

In addition to the closed strings propagating in the bulk, we have
different sectors of open strings, defined by the classes of branes
to which they are attached. A given class of open strings will be
labeled $(p,q)$ if they connect a D$p$-brane and a D$q$-brane. The
relative orientation of the branes is in principle arbitrary, although
we will only consider supersymmetric intersections, i.e. those for
which the $(p,q)$ strings propagating along the intersection submanifold 
have a supersymmetric ground state. 

Each $(p,q)$ system is characterized by a different partition of the
10 space-time dimensions into Neumann--Neumann (NN), 
Dirichlet--Dirichlet (DD),
or mixed (Dirichlet--Neumann (DN) plus Neumann--Dirichlet (ND)): 
\be
10=d_{NN} + d_{DD} + d_{ND} +d_{DN}
,\ee
where $d_{ND} +d_{NN} = p+1$ and $d_{DN} + d_{NN} = q+1$.   
Accordingly, we denote the radii of the torus in these directions by
$R_{NN}, R_{DD}, R_{ND}$ and $R_{DN}$  (some of which could
be infinite). Notice that this labeling of the torus radii depends on
the particular $(p,q)$ system of open strings we focus on. 

The total number of directions with mixed boundary conditions for a
given $(p,q)$ system is  denoted by
\be
\nu \equiv d_{ND} + d_{DN}
,\ee
and for a supersymmetric intersection it must take values
\be
\nu = 0\,({\rm mod} \;4)
.\ee
The simplest case of $\nu=0$ corresponds to parallel branes. Intersections
with $\nu=4$ are all D$p$--D$(p+4)$ systems and their T-duals. Finally,
a prototype of $\nu=8$ system is the D0--D8 intersection and all T-duals. 
So, for example a Type--I model with wrapped D5- and D1-branes contains
closed strings and open strings in all $\nu=0,4,8$ sectors.

We always  assume 
that the system is at weak coupling so that  
the mass of the D-branes is large and perturbation theory 
around the D-brane background is a good approximation. 
In particular in this limit we can neglect 
brane creation in the vacuum and can neglect the effects of 
perturbations of the brane itself on the thermodynamics. 
In later sections we discuss the meaning of this 
assumption in more detail, and in particular the thermodynamic systems 
in which it might be expected to break down.

One additional point. For calculations in purely perturbative 
closed-string theories, an important question was whether 
to take winding number and momentum to be conserved in the compact 
dimensions. Indeed, the thermodynamic properties are typically found to be 
qualitatively different if these quantum numbers are conserved.
In this paper we are ultimately interested in the thermodynamic 
properties of two or more systems in equilibrium in a D-brane background. 
Hence in all of our calculations we {\em do not} conserve winding number
or momentum in the compact dimensions, since  
for example D-branes can absorb and emit 
momentum, and winding number can be transferred from a gas of open 
strings on the brane to a gas a closed strings in the bulk.

\subsection{Single-String Density of States}

The   open strings in the $(p,q)$
sector have NN momenta, DD windings, and oscillators in all transverse
directions. Notice that they {\it do not}
 have momentum or winding quantum
numbers in the ND or DN directions. As a result, the thermodynamic
quantities of the $(p,q)$ system are independent of the ND, DN moduli
$R_{ND}, R_{DN}$.   Using T-duality, we assume that all radii are
larger or equal
than the T-selfdual radius: $R_{NN}, R_{ND}, R_{DN}, R_{DD} 
\geq 1$ in string units.\footnote{We shall use throughout the paper string
units with Regge slope parameter $\alpha'=1$, which has the property that
the T-selfdual
 radius is $R_{\rm self-dual} =1$.} In fact, when the radii are of stringy size,
the NN or DD character is not sharply defined, but then we shall see
that thermodynamics depends only on $\nu$ as a dimensional parameter,
which is  T-duality invariant.

The single-string  energy is given by  
\be
\label{ens}
\varepsilon^2 =  ({\vec p}\,)^2  +\left( 
{\rm Osc}_{\sigma} -a_{\sigma}\right).
\ee
The constant $a_\sigma$ is the normal ordering intercept
with the spin structure $\sigma $: $a=1$ for
bosonic strings, $a_{\rm NS} = 1/2$ for Neveu--Schwarz spin
structure and $a_{\rm R} =0$ for the Ramond sector, 
in the case of superstrings. Open strings on D-branes have
$a_{\rm R} =0$ and $a_{\rm NS} = (\nu-4)/8$.  
Here ${\vec p} $ is the momentum in the  spatial NN  
directions plus the contributions from the open-string windings in the 
DD directions,   
\be 
\label{eq:pn}
 ({\vec p}\,)^2 = 
\sum_{i\in NN} \frac{n_i^2}{R_i^2 }+
\sum_{i\in DD} l_i^2 R_i^2 ,
\ee
where $R_i $ are the torus radii. The oscillator part ${\rm
Osc}_{\sigma}$ receives integer contributions from world-sheet
bosons in NN or DD directions, but  half-integer contributions
from the bosons in ND and DN directions. World-sheet fermions
contribute according to the spin structure (which is correlated with
the bosonic modding, \ie fermions have the same modding as bosons in
the R sector, and opposite in the NS sector). 

The number of states with a particular 
value of the oscillator level, $ {\rm Osc} =n$, is obtained from 
\be 
\label{lden}
d(n)=\frac{1}{2\pi i} \oint {\dd q \over q^{n+1}}
\; f(q)
,\ee
with the function 
\be
f(q) = \Tr_{\rm osc} \,q^{\rm Osc}
\ee
denoting the oscillator trace generating function. 
In a manifestly supersymmetric treatment, such as the Green-Schwarz
formalism, the modding of fermionic and bosonic oscillators must
be the same: integer in NN+DD directions and half-integer in ND+DN
directions. Thus, we get a factor of
\be
\prod_{m=0}^{\infty} \left({1+q^m \over 1-q^m}\right) 
\ee
for each of the $8-\nu$ transverse NN+DD directions, and a factor of 
\be
\prod_{m=0}^{\infty} \left({1+q^{m+\half} \over 1-q^{m+\half}}\right)
\ee
for each of the $\nu$ directions with  ND or DN boundary conditions.
 In addition, we have the degeneracy from
fermionic zero modes, which can be determined from the size of the
corresponding massless multiplets. It is given by $C_{\nu} =
2^{4-\nu/2}$, \ie a ten-dimensional vector multiplet $(C_0 =16)$
 for $\nu=0$,
a half-hypermultiplet $(C_4 = 4)$ for $\nu=4$, and
a single state  for $\nu=8$. These degeneracies may be
affected by `flavour' factors, such as a factor of $N^2$ for $N$
parallel D-branes in $\nu=0$, a factor of $2$ for both orientations  
(\ie $(p,q)$ and $(q,p)$ strings)
 in $\nu\neq 0$ sectors, or factors of $1/2$ due to 
orientifold projections. 

Putting all factors together, and using Jacobi's and Dedekind's 
functions, we find, for each single orientation and flavour sector: 
\be
\label{efeg}
f(q) = \left({\theta_2 (q) \over \eta(q)^3}\right)^{4-{\nu\over 2}}
\,\left({\theta_3 (q) \over \theta_4 (q)} \right)^{\nu \over 2}
.\ee

 A crucial property of (\ref{efeg}) is that the contribution from ND and
DN directions has no modular anomaly under modular transformations,
consistent with the absence of zero modes in those directions.

Using the modular properties of $f(q)$ the integrand of (\ref{lden})
 can be approximated 
for $q\rightarrow 1$~\cite{hardy}. If we define $q=1-\xi $ then 
\be
\frac{f(q)}{q^{n+1}} \sim \left(\frac{\xi}{2\pi}\right)^{4-\nu/2}\, \exp 
\left( 2 \pi^2 /\xi + (n+1)\xi \right).
\ee
There is a saddle point at $\xi_0=\sqrt{2} \pi/\sqrt{n+1} $. 
Upon expanding about this and doing the Gaussian integral 
which results, one finds that the  following level-density, a
generalization of results in   
\cite{others,bowick,deo}, 
\be 
d(n,n_i,l_i)\sim n^{(\nu-11)/4} \;e^{\beta_c
\sqrt{n} },
\ee
where $\beta_c = 2\sqrt{2}\pi$. 
The level density tells us the number of oscillator modes in a 
given momentum and winding sector.  The single-string density
of states is then obtained by summing over all zero-mode quantum
numbers, with the constraint of Eq. (\ref{ens})    
\be
\label{asamp}
\omega (\varepsilon) \dd \varepsilon \sim  
\sum _{l_i,n_i}  
 \varepsilon^{(\nu-9)/2} 
e^{\beta_c \sqrt{n(\varepsilon)}}
\;  \dd\varepsilon,  
\ee
with the function $n(\varepsilon)$ defined by (\ref{ens}) upon setting
${\rm Osc}=n(\varepsilon)$. 
Note that, since all dimensions are taken to be
compact, there are no integrations over continuous momenta and no volume 
factors\footnote{Our separation of quantum numbers between oscillators and
momentum/winding is natural in the context of backgrounds with a clear
geometric interpretation, such as tori or other sigma-models. However, it is
not strictly necessary, and similar results can be obtained for more
general CFT's.}. These are recovered once we let the dimensions become large
and perform a large-$\varepsilon$ expansion of  $n(\varepsilon)$;
\be 
\label{expand}
\sqrt{n(\varepsilon)} = \varepsilon  -
\sum_{i\in NN} \frac{n_i^2}{2\,\varepsilon \,R_i^2 }-
\sum_{i\in DD} \frac{l_i^2 R_i^2 }{2
\,\varepsilon} +\ldots
\ee
The summations over $n_i$ and $l_i$ can be approximated by 
Gaussian integrals when each successive 
term in Eq. (\ref{expand}) is small and the summations are 
nearly continuous, which requires
\ba
\label{comp}
R_{NN}^2 &\gg &   1 /\varepsilon \nn\\
R_{DD}^2 &\ll &   \varepsilon \, .
\ea
The first condition is always satisfied (by assumption, since we 
have T-dualized the small directions and since 
$\varepsilon  \gg 1$
for the asymptotic approximation (\ref{asamp}) to be accurate at all). 
The second condition gives a threshold energy, above which the string is 
energetic enough to wind around Dirichlet directions which we shall 
define (for later convenience) as,
\be 
\varepsilon_0 = 2 R^2_{DD}/\beta_c.
\ee
When $\varepsilon \ll \varepsilon_0$ the Dirichlet directions 
can only contribute when $l_i$ is zero.
The single-string density of states in the limits of high and 
low energy is
\be
\label{omegafirst}
\omega(\varepsilon) \dd \varepsilon  
 = \left\{
\begin{array}{ll}
\beta_c\; V_{NN}
(\beta_c\, \varepsilon) ^{-d_{DD}/2} 
\;e^{\beta_c\, \varepsilon}
\;\dd\varepsilon & \varepsilon \ll \varepsilon_0 \nn\\
\,\nn\\
f \;\beta_c  
\;
e^{\beta_c \,\varepsilon}\;\dd\varepsilon & \varepsilon \gg \varepsilon_0.
\end{array}
\right.
\ee
where $V_{NN}$ is the  
volume of the spatial Neumann directions, and where we have defined 
the ratio of volumes as
\be 
\label{rat}
f = \frac{V_{NN}}{V_{DD}}.
\ee
Note that the density of states changes as we go through the 
energy threshold and strings are able to wind in the Dirichlet 
direction. The exponent in  Eq. (\ref{omegafirst})  counts 
the number of DD directions in which strings are {\it not able}
to wind  
and hence it can be generalized 
to the case where the Dirichlet directions have varying sizes; 
one instead finds a series of thresholds and an $\omega(\varepsilon)$ 
which interpolates between the two extremes in Eq. (\ref{omegafirst}).
To accomodate this possibility, we define 
an effective dimension for open strings $d_o(\varepsilon)$ 
which is the number of DD dimensions around 
which the
strings {\em cannot} wind (\ie which do not obey Eq. (\ref{comp})),
\be 
0 \leq d_o \leq d_{DD}.
\ee
We  define the volume of these dimensions to be
$V_{o}$ and then have 
\be
\label{omega} 
\omega({\varepsilon})= \beta_c\,f \; V_o
 \frac{e^{\beta_c \varepsilon}}
{(\beta_c\,\varepsilon)^{\gamma_o+1}}
,\ee
where the critical exponent $\gamma_o$ is given by 
\be 
\label{gamma}
\gamma_o = \frac{d_o}{2}-1.
\ee
We can, at this point, also define an
{\em effective} number of large space-time dimensions which is 
a function of $\varepsilon$,
\be
\label{deff}
D_o(\varepsilon)=d_{NN}+d_o(\varepsilon) \leq 10-\nu.
\ee
Notice that, in this definition, we have excluded possible `large'
ND+DN directions, as they play no role in opening thresholds.  

\vspace{0.5cm}
\noindent{\it \underline{Examples}}
\vspace{0.5cm}

For what is normally meant by a D$p$-brane (\ie a $(p+1)$-dimensional  
 world-volume 
with $9-p$ non-compact Dirichlet directions and $p$  spatial
non-compact 
Neumann directions), we would recover 
$d_{NN}=p+1$ and always have
 $\varepsilon < \varepsilon_0 = \infty$, 
and hence
\be 
\label{omega2}
\omega(\varepsilon) \,\dd \varepsilon  
= \beta_c \;V_{NN}
\;(\beta_c \,\varepsilon) ^{-(10-d_{NN})/2}
 \;e^{\beta_c \,\varepsilon}\,\dd\varepsilon =
\beta_c \;V_p
\;(\beta_c \,\varepsilon) ^{(p-9)/2} \;e^{\beta_c\, \varepsilon}\dd\varepsilon ,
\ee
in accord with Ref. \cite{us}. However Eq. (\ref{omega}) carries 
useful additional information about what happens to the density of states 
as dimensions expand or contract. In particular we see a kind of 
behaviour that is 
familiar from closed-string thermodynamics. When a Dirichlet 
direction becomes sufficiently small or strings become sufficiently 
energetic
(given by Eq. (\ref{comp})), open strings are able to wind around it. 
This is reflected in the density of states by that 
dimension becoming `compact'; 
when all the Dirichlet directions are compact, 
and $\varepsilon\gg \varepsilon_0 $ we instead find 
\be 
\label{omega2b}
\omega(\varepsilon) \,\dd \varepsilon  
=
\beta_c \;\frac{V_p}{V_{D-p-1}}
\;e^{\beta_c\, \varepsilon}\;\dd\varepsilon .
\ee

Open strings in Type--I theories propagate 
through the whole Neumann
volume, and hence the calculation is the same as for open strings on 
the $(d_{NN}-1)$-brane in Type--II theory; if the space is non-compact
we would again take $\varepsilon\ll\varepsilon_0$ 
in which case we recover the non-compact result of Ref. \cite{dienes} 
when $\gamma_o=(D_o-d_{NN})/2-1=-1$.
However when any of the dimensions are compact, there is an energy dependence
in the density of states in this case as well.   
In particular, we could consider the case where 
there are 3+1 directions which are
much larger than the string scale,
$c$ compactified directions of order the string scale and 
$6-c$ compactified directions which are much smaller 
than the string scale. (After a duality transformation 
these become $6-c$ Dirichlet directions much larger than the string scale
with heavy winding modes.) 
In this case the density of states depends on how many of 
the compact dimensions the strings are able to probe.
That is, when the internal energy is very large the open strings are 
able to propagate throughout the whole space and we have 
$\gamma_o+1=0$. 
However, when $\varepsilon \ll \varepsilon_0$ 
we instead have $\gamma_o+1=d_{DD}/2 = 3-c/2 $ 
(since in the T-dualized theory 
the very small Neumann directions have become Dirichlet directions 
with winding modes which are too heavy to excite). 
For cases of phenomenological interest there may be 
varying radii and hence a complicated energy dependence
in $\gamma_o$. 

\vspace{0.5cm}
\noindent{\it \underline{Free Energy}}
\vspace{0.5cm}

The canonical free energy $F=-{1\over \beta} \,{\rm log}\,Z$
 is now given by the Laplace transform of 
$\omega(\varepsilon)$. The different energy thresholds  in  
(\ref{omegafirst}) translate into analogous thresholds in temperature,
since $\bra \varepsilon\ket \sim (\beta-\beta_c)^{-1}$, we can regard
$V_o$ as the volume of the DD directions satisfying  
$$
1\gg {\beta-\beta_c \over \beta_c} \gg {1\over  R_{DD}^2}  
.$$  
In this region, we  have the approximate behaviour 
\ba
\label{free}
{\rm log}\,Z &\sim& 
\int \dd\varepsilon \, \omega(\varepsilon) 
 \,
e^{-\beta \varepsilon} \nn\\
&=& f
\,V_o\, \left( \frac{\beta-\beta_c}{\beta_c}\right) ^{{d_o \over 2}-1}.
\ea
Notice that, as soon as we assume compact dimensions, we require 
additional information about the system as a whole
(\ie its total energy) in order to calculate the free energy.
When all dimensions are assumed to be non-compact from the outset
this problem of course never 
arises since it would require an infinite amount of energy for 
a string to wind so that the above is always valid.
However, the thermodynamic limit means taking
an infinite volume and filling it with a finite energy density. 
Consequently, 
in a compact space of {\em any} size, there can in principle be enough 
energy available for strings to wind. 

\subsection{Euclidean Approach} 

This question of the effect of compact dimensions 
on the thermodynamic limit 
(which was just stated in rather simple minded terms) 
is of central importance and was addressed for 
closed strings in Ref. \cite{deo}. It will require a proper understanding 
of the microcanonical ensemble which will be the main goal of the next 
section. As a first step, let us recalculate the density of 
states using the method of Refs. ~\cite{greent,mav,us} 
in which the free energy is determined 
by compactifying in an imaginary time direction with periodicity $\beta$.
The particular benefit of this method is that it identifies the leading 
and subleading contributions to the free energy as singularities 
of the partition function in the complex $\beta $ plane, as a result
of `massless' closed-string exchange between the branes.
Our main task therefore is to find the structure 
of these singularities.

In the $(p,q)$ sector the  free energy  
is a sum of terms for different spin structures
$\{\sigma\}$, each one of  the form  
\be
{\rm log}\,Z_{(p,q,\sigma)} = \half \int {dt\over t} \Tr_{\rm open}
\,\,\, \CO_{\rm GSO}^{(\sigma)}\,\,\,e^{- t
\Delta_{\rm open}},
\ee
where $\CO_{\rm GSO}$ is (a piece of) the GSO projector, \ie
$\pm 1$ or $\pm (-1)^F$ depending on the particular spin structure.  
The open-string  world-sheet hamiltonian is 
\be
 \Delta_{\rm open}   =  4\pi^2  n_{\sigma}^2 / 
 \beta^2 + 
({\vec n} / R_{NN})^2
 + (
{\vec l} \, R_{DD})^2  + ({\rm
Osc} -a)_{\sigma}  ,
\ee
with $n_{\sigma}$ an integer or half-integer depending on the bosonic
or fermionic statistics of the space-time states being traced over. 
For notational simplicity, we
 are assuming here equal radii within each NN or DD class of directions.
The generalization to arbitrary radii is straightforward. 

Upon Poisson resummation in $n_{\sigma}$, 
${\vec n}$ and ${\vec l}$ 
we find 
\ba
{\rm log}\,Z_{(p,q,\sigma)} &\sim& \beta
\cdot f \cdot \int
{dt\over t} t^{-(d_{NN} + d_{DD})/2} \,
\Tr_{\rm osc} \,\CO_{\rm GSO}^{(\sigma)} \,e^{-t(
 {\rm Osc} -a)_{\sigma}}  \times
\nn\\
&&
\hspace{6cm}
\sum_{n, {\vec n}, {\vec l}} (-1)^{n{\bf F}_{\sigma}}\,
 e^{-{2\pi^2 \over t}
\Delta^{(0)}_{\rm closed}}, 
\ea
where $n$ is now an integer  
and ${\bf F}$ is the space-time fermion number. The zero-mode action
 now reads   
\be
2\; \Delta^{(0)}_{\rm closed} = 
({\vec n}\,R_{NN})^2
+({\vec l}
/ R_{DD})^2
+ n^2 \beta^2 /4\pi^2 
.\ee
This notation suggests that the appropriate  modular transformation
 to the closed-string channel 
is given by  the change of variables
$
t = 2\pi^2 / s
$, 
which acts on the theta functions from the oscillator traces
in the following universal form:  
\be
\label{tros}
\Tr_{\rm open\,\,osc} \,\CO_{\rm GSO}^{(\sigma)}
 \,e^{-t( {\rm
Osc} -a)_{\sigma}} \sim s^{-4+\nu/2} \,\bra {\rm D}p_{\rm
osc}|  
\CO_{\rm GSO}^{(\sigma)}\,e^{-s(
{\rm Osc}_L + {\rm Osc}_R -a_L -a_R)_{\sigma}} |{\rm D}q_{\rm
osc}\ket.   
\ee
The power of $s^{-4+\nu/2} 
$ in this formula comes from the modular transformation  
of eight-dimensional transverse oscillator traces
as in (\ref{efeg}).  
The closed strings propagating between D-brane boundary states
 (as given for example in \cite{greenbs})  
do not have windings and momenta at the same time in any direction,
so that we can assume trivial level matching  for all spin structures
when evaluating Eq.
 (\ref{tros}):     
\be
({\rm Osc} - a)_L = ({\rm Osc}-a)_R
.\ee

Now, putting everything together, one obtains an expression with the 
 form of a closed-string propagator:  
\be
\label{madre}
{\rm log}\,Z \sim +f \int  
ds\,\sum_{\lambda} g_{\lambda}\, e^{-s\lambda}
,\ee
where  the quantities      
\be
\label{eigen}
2\,\lambda = n^2 \beta^2 /4\pi^2  +({\vec l}
/ R_{DD})^2 
+({\vec n} \;R_{NN} )^2
+4\left({\rm Osc} -a\right)
\ee
are  those eigenvalues of the full 
closed-string kernel, $\Delta_{\rm
closed}$, with non-vanishing  overlap with both D-brane states:
\be
\bra \lambda | {\rm D}p, {\rm D}q\ket \neq 0
\ee
and $g_{\lambda}$ denotes the multiplicity of a given eigenvalue. The
spectrum is discrete at finite volume, which justifies writing the free
energy as a discrete sum. 

This result admits some simple generalizations. In the presence
of orientifold planes, analogous considerations hold, replacing the
D-brane boundary states by orientifold boundary states $|{\rm O}p\ket$,
and introducing the orientation projection in the open-string sector.
The result 
is an expression of the same general form as Eq. (\ref{madre}), with a
different modding of  quantum numbers in the closed-string sector (see
Ref. \cite{mav}).
For example, a crosscap
restricts the winding numbers to be even.

One can also generalize the
previous formulas to the case where various
 D-branes and orientifold planes have
some transverse separation $L_{DD}$. This simply introduces a stretching
energy for the open strings of the form $L_{DD} /2\pi$, which
translates into a new term in the closed-string channel expression Eq.
 (\ref{madre}), an
insertion of 
$
e^{-L^2 / 2 s}
$ 
in the proper-time integral.    

\vspace{0.5cm}
\noindent{\it \underline{Critical Behaviour}}
\vspace{0.5cm}

Formula (\ref{madre})
 is ideally suited for estimating the critical behaviour
of the free energy. At finite volume, singularities appear only when
some eigenvalue vanishes as a function of the temperature, which in turn
can be extracted from the behaviour of the integral (\ref{madre}) for
large proper times  $s\rightarrow \infty$.
A natural ultraviolet cut-off for the $s$-integral corresponds to $s\sim
1$ in string units, 
leading to a representation in terms of an incomplete Gamma function  
\be
\label{fmnu}
{\rm log}\,Z \sim f\,
\sum_{\lambda } g_{\lambda}\;\lambda^{-1}
\;\Gamma\left[1\, ; \,\lambda\right]
,\ee
with the simple scaling     
\be
\label{fmcero}
 {\rm log}\,Z \sim f \,
\sum_{\lambda } {g_{\lambda} \over  \lambda} \;e^{-\lambda} \approx
f\sum_{\lambda<\ord (1)} {g_{\lambda} \over  \lambda} 
.\ee
Notice that the cut-off in the Schwinger parameter at $s\sim 1$ effectively
restricts the eigenvalue sum to $|\lambda|<\ord (1)$ in string units.

From Eq. (\ref{eigen}), we find that $\lambda$ is an increasing function of
$\beta^2$. In all cases of interest, the NS--NS scalar with thermal
winding number $n=\pm 1$~\cite{ian, awit, brandvafa}
survives the GSO projection, and therefore
 has a tadpole on the D-brane state:
\be
\bra {\rm D}p|n=\pm\ket \neq 0
\ee
with degeneracy $g=2$. In this sector, there is a negative Casimir energy
(from $a_{\rm NS} =1/2$) and the   thermal
scalar becomes massless   at the leading (lowest temperature)
 zero of the  eigenvalues  $\lambda(n =\pm 1)$.  
 The origin of the thermal scalar in the closed-string sector is responsible
for the universality of   this singularity: 
\be 
\label{hagt}
\beta_c^2 = 8\pi^2 
,\ee
the standard Hagedorn temperature of Type--II strings.
The most important subleading terms correspond to a family 
of nearby singularities for $R_{DD} \gg 1$
given by
\be
\label{rsing}
\label{temps}
\beta^2_c(l) = \beta^2_c \left(
1-\frac{l^2}{ 2 R_{DD}^2}\right).
\ee
The multiplicity for large $l$  is given by\footnote{Notice that the
multiplicity of any critical eigenvalue is even because $\lambda \propto
\beta^2 n^2$ and $n=0$ states do not produce critical behaviour.}  
\be
g_{l} \simeq 2\, {\rm Vol}({\bf S}^{d_{DD} -1}) \,l^{d_{DD}-1}
.\ee
Notice that the large world-volume of the 
D-brane, $ R_{NN} \gg 1$,
does not introduce any new singularities at `low' temperatures. 

If a given eigenvalue vanishes at $\beta=\beta_{\alpha}$,
\be
\lambda_{\rm critical} = 2n^2\left({\beta_{\alpha}\over \beta_c}\right)^2\,
 \left({\beta-\beta_{\alpha}
\over\beta_{\alpha}}\right) +
\ord\left(\beta-\beta_{\alpha}\right)^2
,\ee
the free energy in the vicinity of
$\beta_{\alpha}$ has a leading singular piece (from the analytic structure
of the incomplete Gamma function)   
\be
\label{sing}
{\rm log}\,Z_{{\rm sing},\alpha} \sim 
f\,\left({\beta-\beta_{\alpha} \over\beta_{\alpha}}\right)^{-1}
.
\ee
The Taylor expansion of the regular part around $\beta=\beta_{\alpha}$ 
 is of some interest for the
calculations in the next section. It can be parametrized in the form
\be
\label{texp}
{\rm log}\,Z_{{\rm reg},\alpha} \sim a_{\alpha} \,V_{NN} -\rho_{\alpha}
\,V_{NN} (\beta-\beta_{\alpha}) + \ord \left( V_{NN}\,(
\beta-\beta_{\alpha} )^2
\right),
\ee 
where $a_{\alpha}$  and $\rho_{\alpha}$
respectively 
have dimensions
 of number and energy  density on
 the world-volume of the intersection. It is convenient
to extract a power of $V_{NN}$ when studying large world-volumes, so that
$a_{\alpha}, \rho_{\alpha} =\ord (1)$ in string units. In
particular, this is also true when
 the transverse volume is also large in string units
$ V_{DD} \gg 1$, since
 both $a_{\alpha}$ and $\rho_{\alpha}$ get most of their
contribution from the sum over non-critical eigenvalues in Eqs.
 (\ref{fmnu}) and (\ref{fmcero}). The eigenvalues
(\ref{eigen}) are densely distributed with spacing $\Delta \lambda \sim
1 /R^2_{DD}$, and the sum over non-critical eigenvalues is itself of
 $\ord (V_{DD})$. Furthermore, although $a_{\alpha}$ and $\rho_{\alpha}$ can be
complex in general,  explicit inspection of Eqs.
 (\ref{fmnu}) and (\ref{fmcero}) shows that the critical density at
the leading Hagedorn singularity $\beta_{\alpha} =\beta_c$ 
is real and positive $\rho_c >0$ in the
same limit, as well as the critical entropy density $a_c >0$
  (one uses the fact that all $\lambda\neq \lambda_0$ are positive at
$\beta_c$,
and that the function   
$$
\lambda^{-1} \,\Gamma \left[1\,;\,\lambda\right]
$$
is positive and monotonically decreasing for $\lambda>0$).         

It is worth emphasizing,
since this will be crucial later on, that
all the quantities which govern the physics are defined
on the world-volume of the intersection. Thus, for an isolated brane,
 Hagedorn behaviour (long string dominance for example)
`switches on' when critical densities are
reached {\em on the brane}.

If the transverse space in DD directions is non-compact 
the singularity structure displayed in Eq. (\ref{sing}) changes. The
singularities  
(\ref{sing}) coalesce with the Hagedorn singularity, changing the
analytic properties of the free energy. Going back to Eq. (\ref{madre})
and converting the sum over ${\vec l}$ into a continuous integral, we
obtain the result of Eq. (\ref{free}) with $d_o = d_{DD}$,  
\be
\label{nonco}
{\rm log}\,Z_{\rm sing} \sim \Gamma(1-d_{DD}/2)\,
 \left(\beta-\beta_c\over\beta_c\right)^{
-1+d_{DD} /2}
,\ee
with a logarithmic correction if $d_{DD}$ is an even integer.
 This is in accord with our previous estimate from 
$\omega(\varepsilon)$ with 
$\varepsilon
\ll \varepsilon_0$. In particular, for $\nu=0$,  we recover the result of
Ref.  ~\cite{us}; $\gamma = (7-p)/2$.  

These results
contain all the required information to pass to the microcanonical ensemble, 
including the dependence on energy thresholds.
For example in a compact space we can simply 
leave the free energy as a sum;
\be
\label{freesum}
{\rm log}\,Z \approx  
f\,\beta_c^2 \sum_{l<\ord (R_{DD})} c_l\,
 {g_l  \over \beta^2-\beta_c(l)^2 }.
\ee
The analytic structure is characterized by a set of isolated
singularities at $\beta = \beta_l$.  

It can be verified that upon taking the inverse Laplace transform 
of Eq. (\ref{freesum}) one recovers the full single-string density of
states, $\omega(\varepsilon)$, of Eq.
 (\ref{omegaopen}).
The relevant contour in $\beta $ may be deformed to give a sum over
poles, so that $\omega(\varepsilon)$ may be written
\be 
\omega(\varepsilon) = \beta_c\,e^{\beta_c\,\varepsilon} f
\sum _l g_l \;e^{-l^2 {\varepsilon /\varepsilon_0}}  .
\ee
Only the first term contributes when $\varepsilon \gg \varepsilon_0$.
When $\varepsilon \ll \varepsilon_0$ we can instead approximate 
the sum over $l$ by an integral which gives a factor 
$(\varepsilon/\varepsilon_0)^{-d_{DD}/2}=  V_{DD}
(\beta_c\varepsilon )^{-d_{DD}/2}$ as required. At intermediate energies we 
recover the full energy-dependent effective dimensions of Eq.
(\ref{omega}).

\subsection{Random Walks}
In this subsection we rederive the previous results on the
single-string density of states from the heuristic random walk picture
of a highly excited string (see for example \cite{pols}).
   In addition to providing a nice physical
interpretation and checks of the calculations, this point of view
leads 
to some possible generalizations beyond toroidal backgrounds.

As a warm-up, we derive the distribution function $\omega (\ep)$ for
closed strings in  $D$ large 
space-time dimensions. The energy $\ep$ of  the
string is proportional to the length of the random walk. The number
of walks with a fixed starting point 
 and a given length $\ep$ grows exponentialy as $\exp
\,(\beta_c\,\ep)$.\footnote{The proportionality constant $\beta_c$
 depends on
the bulk details of the string, such as  the
presence of fermions on the world-sheet, but it is independent of 
boundary effects (\ie open or closed strings).}  
 Since the walk must be closed, this overcounts by a
factor of the volume of the walk, which we shall denote by 
 $V({\rm walk}) = W$. Finally, there is a
factor of $V_{D-1}$ from the translational zero mode, and a factor of
$1/\ep$ because any point in the closed string can be a starting point.
The final result is
\be
\label{rw}
\omega(\ep)_{\rm closed} \sim V_{D-1}\cdot {1\over \ep} \cdot
{e^{\beta_c\,\ep} \over W}
.
\ee
Now, the volume of the walk is proportional to $\ep^{(D-1)/2}$ if it is
well-contained in the volume $(R\gg \sqrt{\ep})$, or roughly $V_{D-1}$
if it is space-filling $(R\ll \sqrt{\ep})$. From here we get the
standard result \cite{others,bowick,brandvafa,deo}. We  have
\be
\omega (\ep)_{\rm closed}
 \sim V_{D-1} \,{e^{\beta_c \,\ep} \over \ep^{(D+1)/2}}
\ee
in $D$ effectively non-compact space-time dimensions, and 
\be
\omega (\ep)_{\rm closed} \sim {e^{\beta_c \,\ep} \over \ep} 
\ee
in an effectively compact space.     

We can generalize this analysis to open strings in a general $(p,q)$
sector by a slight modification of the combinatorics. The leading
exponential degeneracy of a random walk of length $\ep$ with a fixed
starting point in say the D$p$-brane is the same as for closed
strings: $\exp (\beta_c\,\ep)$.
Fixing also the end-point at a {\it particular} point of the D$q$-brane
requires the factor $1/W$ to cancel the overcounting, just as in the
closed string case. Now, both end-points  move freely   
in the part of each brane occupied by the walk. This gives
a further degeneracy factor
\be
(W_{NN} \,W_{ND})\cdot (W_{NN} \,W_{DN})
\ee
from the positions of the end-points.    
Finally, the 
overall translation of the walk in the excluded NN volume gives a
factor
$V_{NN} / W_{NN}$. The final result is: 
\be 
\omega (\ep)_{\rm open} \sim {V_{NN}\over W_{NN}} \cdot
 W_{NN+ND} \cdot W_{NN+DN} \cdot {1\over W} \cdot \exp\,(\beta_c\,\ep) \sim  
 {V_{NN} \over W_{DD}} \;   
\exp\,(\beta_c\,\ep) 
.\ee 
Thus, we find that the density of states is only sensitive to the 
effective volume of the random walk in DD directions. If the walk 
is well-contained in DD directions $(R_{DD} \gg \sqrt{\ep})$,  
we find $W_{DD} \sim \ep^{d_{DD}/2} $ and  
\be
\omega (\ep)_{\rm open} \sim {V_{NN} \over  
\ep^{d_{DD} /2}} \;\exp\,(\beta_c\,\ep) 
.
\ee 
On the other hand, if  it is space-filling in DD directions $(R_{DD} \ll
\sqrt{\ep})$, the DD-volume of the walk is just $W_{DD} \sim 
V_{DD}$ and we find  
\be 
\omega(\ep)_{\rm open} \sim {V_{NN} \over V_{DD}}  
\;\exp\,(\beta_c \,\ep) 
,
\ee 
in  agreement with   
 (\ref{omegafirst}) and (\ref{omega}).

The random walk picture gives a geometric rationale for the 
                similarity between non-compact closed-string
and open-string densities of states. It is related to the fact 
that the random walk must `close on itself' in some effective
co-dimension (the full space for closed strings and the DD space
for open strings).  

A further interesting aspect of the random walk derivation is that
it naively generalizes to arbitrary backgrounds. In principle, one
could take for example a group manifold without non-contractible cycles,
showing that it is `available volume', rather than `winding modes' that
really determines the physics of the Hagedorn ensembles. For toroidal
backgrounds these two features cannot be disentangled, and 
 the Poisson resummation performed in the previous section
leads to a nice interpretation of the relevant singularities as
associated to winding modes in the open-string sector. Since we lack
exact results in other backgrounds, we shall continue working with the
language appropriate for toroidal backgrounds,  refering to the
opening of the winding thresholds as synonymous with the general  
`volume saturation' property described in this section.

\subsection{Summary}

We complete this section by collecting the expressions for the 
densities of states of single strings. For the open strings we found
\be
\label{omegaopen} 
\omega({\varepsilon})= \beta_c\,f \; V_o
 \frac{e^{\beta_c \,\varepsilon}}
{(\beta_c\,\varepsilon)^{\gamma_o+1}}
,\ee
where
\be 
\label{gamma2}
\gamma_o =  {d_o \over 2} -1 = 
\frac{D_o-d_{NN}}{2}-1,
\ee
and $D_o$ is the effective number of large dimensions we defined above
(\ie the total number of NN+DD dimensions 
minus the dimensions in which open 
strings have sufficient energy to wind).

The analytic structure of the free energy at the Hagedorn singularity
is given by 
\be
\label{si}
{\rm log}\,Z_{\rm sing} 
 \sim \left\{
\begin{array}{ll}
\Gamma(-\gamma)\;f\;(\beta-\beta_c 
)^{\gamma}
\;,\;\;\;\;\;& \gamma\notin {\bf Z}^+ \cup \{0\}  \nn\\
\;\nn\\ 
{(-1)^{\gamma+1} \over \Gamma(\gamma+1)} \;f\; (\beta-\beta_c
)^{\gamma} \;{\rm
log}\,(\beta-\beta_c) \;,\;\;\;\;\;&\gamma\in {\bf
Z}^+ \cup \{0\}. 
\end{array}
\right.
\ee
with the critical exponent $\gamma=-1$ for compact DD directions.
 If a number $d_{\infty}$ of
DD dimensions are 
strictly non-compact, the form (\ref{si}) is still valid with the replacements
$\gamma\rightarrow \gamma+d_{\infty}/2$ and $f\rightarrow f\cdot V_{\infty}$.

The density of states in the closed-string sector has already been calculated 
 in the context of weakly-coupled strings~\cite{others,bowick,brandvafa,deo}. 
For this we need to define another energy-dependent effective 
space-time dimension, $D_c$, which is the total number of dimensions
minus the number of dimensions around which closed strings
can wind (given by the equivalent of Eq. (\ref{comp})). 
If we also define the volume of this dimension, $V_c$, we then have
\be
\label{omegaclosed}
\omega(\varepsilon)  
 = \beta_c\, V_c
 \; \frac{e^{\beta_c\,\varepsilon}}{(\beta_c\,\varepsilon)^{\gamma_c+1}},
\ee
where 
\be
\gamma_c = \frac{D_c-1}{2}
\ee
is the $\varepsilon$-dependent critical exponent for the closed
 strings. 

According to \cite{deo}, the analytic structure of the {\it partition function}
 at finite volume is given
by a set of poles of even  multiplicity $g_{\alpha} 
=2k_{\alpha}$:
\be
\label{ploes}
Z_{\rm sing, cl} \sim \prod_{\alpha} \left({\beta_{\alpha} \over
\beta-\beta_{\alpha}}\right)^{k_{\alpha}}
,\ee
with $k_{\alpha} =k_c =1$ for the leading Hagedorn singularity
$\beta_{\alpha} =\beta_c$. 
If $D_{\infty}$ space-time dimensions are non-compact, one obtains the same
expression as in (\ref{si}) with $\gamma\rightarrow (D_{\infty}-1)/2$ and
$f\rightarrow V_{\infty}$.    
 
\section{Thermodynamic Properties from the Microcanonical Ensemble}

We now progress to a discussion of the thermodynamic properties of 
isolated and coupled systems in the Hagedorn regime. 
The first part of this section will be mostly taxonomy; we shall 
categorize the systems according to the classical 
(non-stringy) analysis of Carlitz~\cite{carlitz} into those 
for which the Hagedorn temperature is 
a limiting temperature in the canonical (Gibbs) approach, 
and those for which the temperature is non-limiting. 
The non-limiting systems should be further analyzed since it is 
these systems for which the canonical and microcanonical ensembles 
are found to be inequivalent. Indeed in general (although not, 
as it turns out, for most of the cases we shall be examining here) the 
microcanonical ensemble can have regions of negative specific heat, 
indicating the possible onset of some phase transition. 

Clearly then, in order to discuss the thermodynamic behaviour of 
string gases, we ought to work in
the microcanonical ensemble and this is the subject of the second part 
of this section. For this we shall have to tackle 
some additional, purely stringy aspects of the thermodynamics.
The most important question, as we have already mentioned,
is how to take the thermodynamic 
limit when the space has 
compact but large dimensions~\cite{deo}. 
The thermodynamic limit involves letting the volume 
go to infinity whilst keeping the energy density finite. The analysis 
of Ref. \cite{deo} shows that for closed strings there is a rather 
peculiar dependence on dimension; when there are more than two
 space dimensions 
which are just large rather than non-compact, taking the thermodynamic limit
always results in winding modes. Consequently the thermodynamic properties 
depend on whether one assumes space to be compact and supporting 
winding modes, or assumes space to be non-compact from the outset (see however
\cite{osorio}).
For open strings attached to D-branes we shall see that
the issue is further complicated by the fact that the strings can only wind 
in the subspace with DD boundary conditions. 
We shall find an additional parameter, the ratio of NN to DD volumes, 
which plays a central role, determining the effects of winding modes. 

Various results for the different brane backgrounds obtained in both the
canonical and microcanonical ensembles   are tabulated.   
Table (1) shows the dependence on $(\beta-\beta_c) $ 
of the internal energy, $E$,  and the pressure, $P$, 
for both cases, where $\gamma$ stands for $\gamma_o$ and $\gamma_c$ 
(the entries are the values of $X$ in $(\beta-\beta_c)^{X} $ whenever the
thermodynamic function diverges as a power, and denote the value of the
function itself otherwise, \ie logarithmic for $X=0$ and constant for $X>0$). 
Thus, for example, the pressure for open strings scales with temperature 
at a fixed volume in the same way as the free energy; 
\be 
P\sim  \left({\beta-\beta_c \over \beta_c}\right)^{\gamma_o} =  
\left({\beta-\beta_c \over \beta_c}\right)^{(D_o-2-d_{NN})/2}.
\ee
We stress that the table is written   for the thermodynamic limit where 
$D_o$ and $D_c$ dimensions are 
{\em non-compact} with the rest of the dimensions being string scale.
Thus, no effects of winding modes are reflected in the table.  
The leftarrows in the table indicate that the non-stringy
microcanonical ensemble has the same internal energy as the canonical. 
In addition
the microcanonical ensemble expression for $\gamma > 1$ is 
valid at high internal energies~\cite{carlitz}; approximately
\be 
\beta_c\,E \gg 1.
\ee
At lower energies the microcanonical and canonical ensembles coincide. 
Note that parameter $a$ appearing in Eq. (\ref{eq:hag}) is given by 
\be 
a=-\gamma_{o,c} -\frac{D_{o,c}+1}{2}.
\ee
The difference between open and closed strings can be understood
as being due to  
integrating $\rho(m)$ only over dimensions in which the centre of mass 
of the strings can propagate~\cite{us}. This dimension can be different 
for open and closed strings as we saw in the preceding section.

\begin{table}[ht]
{\footnotesize 
\centerline{\begin{tabular}{|c|c|c|c|c|c|}
\hline 
$\gamma$ & \mbox{open} & \mbox{closed} & $P\sim (\beta-\beta_c)^{X} $ 
 & $E_{c} $ 
 & $E_{mc}$  \\
\hline
\hline
&&&&& \\
 $\gamma<0 $  & $d_{NN}>D_o-2 $ & $D_c<1$ &
 $\gamma$ & $\gamma-1$ 
& $\leftarrow $ \\
$\gamma=0 $ & $d_{NN}=D_o-2 $ & $D_c=1$ & $\log $     
& $\gamma-1$ & $\leftarrow $ \\
$0<\gamma<1 $ & $D_o-2 > d_{NN}>D_o-4 $ & $1<D_c<3$ 
& constant & $\gamma-1$ & $\leftarrow $ \\
$\gamma=1 $ & $d_{NN}=D_o-4 $ & $D_c=3$ & constant & $\log $
& $\leftarrow $  \\
$\gamma>1 $ & $d_{NN}<D_o-4 $ & $D_c>3$ & constant & constant
& $-1$  \\
&&&&& \\ \hline 
\end{tabular}
}
}
\caption{Thermodynamic regimes for open and closed strings with $D_o$
and $D_c$ {\em non-compact} space-time dimensions. The remaining 
dimensions are string scale.}
\label{table1}
\end{table}

According to the table, all systems which have 
$\gamma\leq 1 $ are unable to reach the Hagedorn temperature 
since they require an infinite amount of energy to do so. In these cases 
the Hagedorn temperature is limiting, and this is true for 
all open strings with $d_{NN}\geq D_o-4$.
In addition, from this table, one might conclude that 
the Hagedorn temperature is non-limiting for the closed strings 
in any realistic model (\ie one which has $D_c \geq 4 $).
When we place a system with $\gamma\leq 1$ in a heat 
bath, it will however behave as a normal gas in the sense that it is 
able to come to equilibrium with it. 

On the other hand, when 
$\gamma>1 $, the canonical and microcanonical expressions 
disagree at high internal energies. This is an indication that there are large 
fluctuations in the microcanonical system. For these systems, Carlitz 
estimates 
\be 
\label{carl}
E\approx \frac{\gamma+1}{ (\beta_c-\beta)},
\ee
and so we see that the specific heat is negative. 
In addition the energy itself is negative below $T_c$. Thus in these cases
it is difficult to ascertain what is going on as we approach the Hagedorn 
temperature from below, and this is where stringy effects 
become important. 

In the original Carlitz analysis, it was suggested that the temperature 
rises {\em above} the Hagedorn 
temperature at some intermediate energy and approaches it from above. 
However, for stringy systems at least, 
the picture must be very  different  
since the string ensemble in contact with a heat bath above 
$T_c$ is not well-defined~\cite{bowick}. 
Systems which are non-limiting can be taken 
up to $T_c$ by a heat bath but, in the case of an ensemble of strings
in a non-compact space, 
additional energy simply goes into one very long string. 
In either picture of course the broad conclusion is the same -- 
the non-limiting string ensembles cannot come into equilibrium, and 
conventional thermodynamics breaks down near $T_c$. 
However when, as here, we wish to consider two or more such systems 
in thermal contact, a detailed understanding is required.

The problems in defining a reasonable thermodynamics of non-limiting
systems can be bypassed by working in finite volume. In fact, string
interactions contain gravity, which necessarily ruins any thermodynamic
limit due to the Jeans instability. Thus, consistency also demands that
we work in a sufficiently small volume, which still could be large enough
to admit an approximate thermodynamic  description.  
 We shall address the
question of string interactions and their effects on the spectrum
of the theory in the last section of the paper. For the time being,
we shall work in finite volume and investigate to what extent purely
perturbative 
stringy effects affect the definition of the thermodynamic limit.

For weakly-coupled closed-string theories, 
a more complete physical picture in this vein was 
provided by Deo \etal~\cite{deo} who pointed out the pitfalls in taking the 
thermodynamic limit.   
 The main argument of Ref. \cite{deo} 
can be understood as follows. Consider attempting to 
recover table (1) by  
 letting $D$ space-time dimensions become much larger than the string scale.
(We stress that $D$ is not to be confused with $D_o$ or $D_c$; 
it is not a function of string energy $\varepsilon$.)
By giving the large dimensions a common radius $R\gg 1$ whilst
keeping the density $\rho$ constant, we might expect to obtain 
behaviour consistent with $D_c=D$ in the table. 
However, on letting the radius become large, the energy scales as $R^{D-1}$,
whereas the (sufficient) condition that there should be 
no windings in the $D-1$ large 
dimensions is that the total energy obeys 
$E \ll \varepsilon_0 = 2 R^2/\beta_c$. 
Clearly this condition is always violated when $D>3$ in the infinite volume 
limit, and it seems that the larger the radius becomes the more energy there 
is to excite winding modes. This argument indicates  that 
we should instead find behaviour consistent with 
$D_c=1$ in a large but compact space with $D>3$ space-time dimensions.
(This is why we  stressed that the table is only correct 
for $D_o$ and $D_c$ non-compact effective dimensions.)

 What goes wrong
with the naive expectation for the above example of closed strings, 
is that it gives the {\em wrong value of $\gamma$}, 
which depends on the macroscopic
properties of the system such as windings.
These heuristic arguments can only be confirmed
by looking at the multiple string density of states where, for example, we
can find the energy distribution and count the number of winding modes.

So, it is clear that there are two
important issues that we should now address. The first,
which we tackle in the following subsection,
is how to calculate the density of states for a given value of $\gamma$.
A   method based on analytic continuation to complex temperatures
was developed in Ref. \cite{deo} in order
to find the density of states
of closed strings. We shall extend this method to include D-branes
by introducing two approximations (the saddle point and
the branch-cut approximations) which can be used in different
volume and energy limits. 

The second issue, which is more subtle, is of course the correct
value of $\gamma$
for particular physical systems in particular limits.
This is discussed in subsections 3.2 and 3.3 where we
consider examples of various different closed- and open-string
systems. In particular in 3.3
we study open strings in a finite box (equivalent to $\gamma=-1$)
where we can also find a more complete expression for the density 
of states (\ie one which is valid for any choice of volumes and 
energies).

\subsection{Complex Temperature Formalism} 

In order to have a well-defined system, we consider a finite-volume
nine-torus
with $D-1$ large space dimensions of radius $R_{\rm large}\gg 1$,
 and $10-D$ remaining
spatial dimensions of string-scale size $R_{\rm small} \sim \ord (1)$.
 The total volume is then
 \be
V_{D-1} \sim \prod_{\rm large} R_{\rm large}. 
\ee
In the presence of intersecting D-brane configurations, the string Hilbert
space splits into different sectors: closed strings propagating in the
full torus of volume $V_{D-1}$, and open-string sectors characterized by
the number of ND$+$DN directions, the T-duality invariant index $\nu$, in
addition to the world-volume dimensionality $d_{NN}$, and the remaining
$d_{DD}$ DD directions.  
For each open-string sector, we denote by $d_{\parallel} \leq d_{NN}-1$
the number of spatial NN directions, of size $R_{\parallel}$,
 which are also {\it large}, and likewise
 $d_{\perp}
\leq d_{DD}$ the number of {\it large} DD directions of size
$R_{\perp}$. One always has
\be
d_{\parallel} + d_{\perp} = D-1-\nu 
.\ee
It is important to keep in mind that each open-string sector has
a particular factorization of the total volume in NN and DD components.

On general grounds, the microcanonical
 density of states of a gas of strings with
 total energy $E$ can be obtained as an  
  inverse Laplace transform of $Z(\beta)$ (see Ref. \cite{bowick}
for
a review);
\be
\Omega(E) = \int_{C_{\infty}}
 {\dd \beta \over 2\pi i} \, e^{\beta E} \, Z(\beta)
,\ee
with the contour $C_{\infty}$ encircling $\beta=\infty$ clockwise.
 Given the analytic structure
of the canonical partition function as explained in the previous section,
we can estimate $\Omega(E)$ in a high-energy expansion by contour deformation
through the singularities of $Z(\beta)$. The original contour $C_{\infty}$ 
can be split
into pieces running close to the singularities (encircling counter-clockwise
the singularities
if they are isolated, as in figure (1), or the cuts if they are branch-points),
 $C_{\infty} =\cup_{\alpha}
 C(\beta_{\alpha})$.

In the ideal-gas or one-loop approximation the total density of states
factorizes in sectors:
\be
\Omega = \Omega_{\rm closed} \cdot \prod_{(p,q)} \Omega_{{\rm D}p-{\rm
D}q} + 2\;{\rm loop} 
.\ee
  In a given  sector, the     density of states is then written
as a sum $\Omega = \sum_{\alpha} \Omega_{\alpha}$, each term being 
dominated by
the behaviour of $Z(\beta)$ near the singularity $\beta_{\alpha}$     
\be
\label{clsing}
\Omega_{\alpha} \approx e^{\beta_{\alpha}
 \,E + a_{\alpha}\, \Vp} \; \int_{C_{\alpha}} {\dd\beta\over
2\pi i} \; e^{(\beta-\beta_{\alpha})(E-\rho_{\alpha}
 \,\Vp) } \;
Z_{{\rm sing},\alpha}
\ee
where we have used the Taylor expansion (\ref{texp}) of the regular
part of the free energy ${\rm log}\,Z_{{\rm reg},\alpha}$ to
  leading order\footnote{When considering the closed-string sector, all
volume factors in the formulas apply with the replacements  
$\Vp\rightarrow V_{D-1}$ and $\Vt\rightarrow 1$.} 
\be
\label{rego}
{\rm log}\,Z_{{\rm reg},\alpha} \sim a_{\alpha}\,V_{\parallel} -\rho_{\alpha}
 \,\Vp
\,(\beta-\beta_{\alpha}) + \ord \left(\Vp\,(\beta-\beta_{\alpha})^2\right).
\ee
The quantity 
$\rho_{\alpha}\sim \ord (1)$ in string units, is a critical Hagedorn density
on the relevant volume for each sector. The smallness of the
 neglected higher order terms
in (\ref{rego}) must be checked in each situation.  

The singular part for open-string systems  takes the form
\be
\label{critfl}
{\rm log}\,Z_{{\rm sing},\alpha} \sim 
\half\, C_{\gamma}\,g_{\alpha}\,f\,\left(\beta-\beta_{\alpha}
\right)^{\gamma} \; 
\left[{\rm log}\,\left(\beta-\beta_{\alpha} 
\right)\right]^{\delta}
,\ee
with $g_{\alpha}$ is  the degeneracy of the critical eigenvalue
$\lambda_{\alpha}$, $\delta =1$ if $\gamma$ is a positive (or zero)
 integer, and $\delta=0$
otherwise. The constant
\be
\label{cdef}
C_{\gamma} \propto  
\left\{
\begin{array}{ll}
\Gamma(-\gamma)
\;,\;\;\;\;\;& \gamma\notin {\bf Z}^+ \cup \{0\}  \nn\\
\,\nn\\
{(-1)^{\gamma+1} \over \Gamma(\gamma+1)}  
\;,\;\;\;\;\;&\gamma\in {\bf
Z}^+ \cup \{0\}.
\end{array}
\right.
\ee
The volume factor 
\be
\label{efes}
f= 
{\Vp\over V_{\perp}}   
,\ee
and  the critical exponent
$ 
\gamma= 
-1
$. Formula (\ref{critfl}) may also be used for closed strings, where
$\gamma=0, \delta=1$, and the absolute
 normalization is obtained setting $f=1$.

\vspace{0.5cm}
\noindent{\it \underline{The Dominance Rule of the Hagedorn Singularity}}
\vspace{0.5cm}

           The condition
that the full density of states is dominated by the contribution
 from the Hagedorn
singularity  $\beta_0 =\beta_c =2\pi \sqrt{2}$ is 
\be
\label{singdom}
{\rm log}\,(\Omega_c /\Omega_1) \gg 1
,\ee
with $\Omega_1$ the contribution to the density of states from the
 closest singularity to $\beta_c$, which we shall denote $\beta_1$. In all the
situations considered in the present work, $\beta_1$ is real and $\beta_1
<\beta_c$.  
   A necessary
condition is obtained from the leading exponential behaviour of (\ref{clsing})  
\be
\beta_c \,E + a_c \,\Vp \gg \beta_1 \,E + {\rm Re} (a_1)\,\Vp  
\ee
or, estimating
 the coefficient ${\rm Re} (a_1) \approx a_0 + \rho_c' \,(\beta_c -
 \beta_1)$ in a
 Taylor
expansion around $\beta_c$, with $\rho_c' =\ord (1)$ in string units,  
\be
\label{dencon}
 (\beta_c -\beta_1)(E-\rho_c' \,\Vp) \gg 1
.\ee
 This is a necessary condition, albeit not sufficient, as we will learn 
later on in   
 this section.
 Again note that this is a condition
involving the energy density on the brane (or intersection).

If the next-to-leading singularity $\beta_1$ 
 is independent of the T-moduli, then
$\beta_c -\beta_1 \sim \ord (1)$ and condition (\ref{dencon})  reduces
to the requirement of large energy densities in the relevant world-volume
($V_{D-1}$ for closed strings or $\Vp$ for open strings):
\be
\rho \equiv {E\over \Vp} \gg \ord (1) 
\ee
in string units. We shall always work in this regime.

 However,
in many cases $\beta_1$ depends on the T-moduli. For example, 
if some  $R_{\perp} \gg 1$ in open-string systems,
 we have $\beta_c -\beta_1 \sim 1/R_{\perp}^2$ as in
(\ref{rsing}), and
(\ref{dencon})  depends non-trivially on the DD moduli. If $R_{\perp}$
is sufficiently large for a given total energy, the condition (\ref{dencon})
is violated. In this case, 
 it is a better approximation to evaluate the density of states in the
non-compact limit
 $R_{\perp}\sim \infty$, which corresponds to the radius-dependent
singularities coalescing with the Hagedorn singularity, changing the critical
exponent   according to
\be
\label{critinf}
\gamma\rightarrow \gamma_{\infty} =\gamma + {d_{\infty} \over 2} 
,\ee  
with $d_{\infty}$ the number of such `non-compact' dimensions. The volume
factor in (\ref{critfl}) also changes by removing from $V_{\perp}$ the
volume of the `non-compact' dimensions. For example, if all $d_{\perp}$
large DD directions are to be approximated as non-compact, we have  
\be
\label{efeinf}
f\rightarrow f_{\infty} \sim \Vp  
.\ee  
 Thus, we handle the violation of (\ref{dencon}) by a change in the
 critical exponent $\gamma$. After this, the remaining closest singularity is
independent of the DD radius and is separated a 
distance of $\ord (1)$ in string
units.

In the context of these approximation techniques, we shall discuss the
evaluation of the basic integral (\ref{clsing}) for a general value of
the critical exponent $\gamma$. A change of variables  
 gives the following expression for the Hagedorn singularity contribution:
\be
\label{zetav}
\Omega_c \approx e^{\beta_c \,E+a_c \,\Vp} \cdot \omega_{
\gamma}  
\cdot \int_{C(\beta_c)} {\dd z\over 2\pi i} \,
{\rm exp}\left\{x_{\gamma} \,(z+C_{\gamma} \,z^{\gamma} \,[{\rm log}\,(z\,
\omega_{
\gamma})]^{\delta})\right\}
,\ee
where 
\be
\omega_{\gamma}\equiv \left({x_{\gamma} \over f}
\right)^{1/\gamma} 
\qquad
z={\beta-\beta_c\over\omega_\gamma
}\ee
 and  the parameter $x_{
\gamma}$ is  defined by\footnote{The case $\gamma=1$ is excluded from the
present discussion and will be dealt
with separately below.
}    
\be
\label{dex}
x_{\gamma} \equiv f\,\left({E-\rho_c\,\Vp \over f}\right)^{\gamma\over
\gamma-1} = f\,[V_{\perp}\,(\rho-
\rho_c)]^{\gamma\over \gamma-1}
.\ee
 It is  important to keep in mind that, while 
$\beta_c$ is a universal constant for all open and closed sectors,
the `ground state degeneracy' $a_c$, and the critical energy
density $\rho_c$ are sector-dependent, both dimensionally and
numerically.

The representation (\ref{zetav}) of the Hagedorn singularity contribution
is well-suited for the presentation of the two most useful approximation
techniques
in the evaluation of $\Omega(E)$, which we now discuss.  

\vspace{0.5cm}
\noindent{\it \underline{Saddle Point}}
\vspace{0.5cm}

Written in     
 the form (\ref{zetav}), we see that the saddle-point approximation
is good if $x_{\gamma} \gg 1$, provided the neglected terms in the
Taylor expansion of the regular free energy, of order
\be
\Vp \,(\beta-\beta_c)^{2+n} \sim \Vp \cdot {(x_{\gamma})^{2+n} 
\over (E-\rho_c\,\Vp)^{2+n}} \,z^{2+n}
\ee
are small. These terms produce a negligible shift of the saddle point
at $z=z_s \sim \ord (1)$ if 
\be
\label{vil}
{\Vp\; (x_{\gamma})^{1+n} \over (E-\rho_c\,\Vp)^{2+n}} \sim V_{\perp} \,
[V_{\perp}\,(\rho-
\rho_c)]^{2+n-
\gamma \over \gamma-1} \ll 1.
\ee
So, given that we are interested in Hagedorn densities in the world-volume
$\rho > \rho_c$, 
and $V_{\perp} \geq 1$  by construction,
we see that, as soon as $\gamma >1$, 
 (\ref{vil}) is violated at sufficiently high order in the
Taylor expansion. Thus,
 the saddle-point approximation is not applicable to the full
integral for $\gamma >1$, even if $x_{\gamma}\gg 1$.  

On the other hand, for $\gamma <1$  and $x_{\gamma}\gg 1$,
the analytic corrections are under
control and the saddle-point approximation is good,
leading to a Hagedorn-dominated density of states  
 of the form
\be
\label{sa}
\Omega(\gamma<1)_{\rm
 saddle} \approx {\omega_{\gamma} 
 \; e^{
\beta_c\,E +a_c\,\Vp  
+{\gamma-1\over \gamma} \,
z_s \,x_{\gamma} +\Delta_s} \over  \sqrt{x_{\gamma}} }
\;\left[1+\ord \left({\Delta_s\over x_{\gamma}}\right) + 
\ord \left({1\over
x_{\gamma}}\right)\right]
\ee
for $\gamma\neq 0$. The quantity $\Delta_s$ in the exponent is of
order 
\be
\label{deltas}
\Delta_s \sim \Vp\,(\beta_s -\beta_c)^2 \sim {(x_{\gamma})^2 \over \Vp
(\rho-\rho_c)^2}
\sim {\Vp \over [\Vt (\rho-\rho_c)]^{2-\gamma\over 1-\gamma}}
,\ee
and accounts for the small shift (\ref{vil}) of the saddle-point
at $\beta_s$ by the neglected analytic corrections to the free energy.
 The first correction term in square brackets  comes
from the effect of these analytic terms in the evaluation of the
saddle-point integral, while
the second one  is a  `two-loop' correction term,   
controlled by the small parameter $1/x_{\gamma}$.   

The saddle-point approximation leads to the equivalence with the
canonical ensemble, so that
 these results are compatible with the contents of table (1), where
$\gamma=1$ is the maximum critical exponent admitting a canonical
analysis. 

By explicit inspection of the saddle-point equation, one can check that
the dominant saddle point (the one with largest real part) for all
values of $\gamma<1$ of the form (\ref{critinf}) has $z_s$ real and
positive, so that 
 there are large exponential contributions to
$\Omega_c$ for $\gamma<0$ 
 in the saddle-point, $x_{\gamma} \gg 1$, limit.   This
implies a refinement of the condition (\ref{dencon}). The contribution
from the next-to-leading singularity $\beta=\beta_1$ to the entropy
is also linear in $x_{\gamma}$ and it differs from the Hagedorn one
by the replacement
\be
x_{\gamma}\rightarrow (g_1)^{1/1-\gamma} \,x_{\gamma}
,\ee
with $g_1$ the multiplicity of the singularity at $\beta_1$.
For the DD-moduli-dependent singularities (\ref{freesum}) we have
$g_1 >1$ and the condition
(\ref{dencon}) must be refined to  
\be    
\label{Hdr}
(\beta_c -\beta_1)(E-\rho_c' \,\Vp) \gg
\left\{
\begin{array}{ll}
{\rm max}\,(1, x_{\gamma})
\;\;\;\;\;\;&{\rm if}\;\;\;\;{\gamma<0}
  \nn\\
1\;\;\;\;\;\;&{\rm if}\;\;\;\;{\gamma\geq 0.}
\end{array}
\right.
\ee 
We shall refer to this condition as the Hagedorn-dominance-rule (H.d.r.).
Notice that it is  equivalent to the requirement that the
saddle-point (canonical)
 temperature be `close' to the Hagedorn temperature:
$|\beta_c -\beta_s | \ll |\beta_c -\beta_1 |$.  If condition
(\ref{Hdr}) is violated, one has to change the critical exponent
$\gamma$ according to (\ref{critinf}) and the conditions for 
single-singularity dominance in the new regime must be considered.

If $\gamma=0$, the `loop-expansion parameter' $x_0 =
x_{\gamma=0} =f$,
independent of the energy, and we find  
\be
\label{sacero}
\Omega (\gamma=0)_{\rm saddle} \approx   
{e^{\beta_c\,E+a_c\,\Vp +f +\Delta_s} \over \sqrt{f}}\;\left( 
{ E-\rho_c\,\Vp \over f} \right)^{f-1} \;\left[1+\ord \left({\Delta_s
\over   
x_0}\right) + \ord \left({1\over x_0}\right)\right]  
,\ee 
with $\Delta_s$ as in (\ref{deltas}) with $\gamma=0$.  

All the systems with $\gamma<1$, whose behaviour is well approximated
by a saddle point, $x_{\gamma} \gg 1$,  show canonical thermodynamic
 behaviour, as in table (1). For these systems the internal energy
diverges at the  Hagedorn temperature, which can be considered as
limiting. We shall denote these systems as ${\bf L}[\gamma]$.  
          
\vspace{0.5cm}
\noindent{\it \underline{No Saddle Point}}
\vspace{0.5cm}

If the conditions for the saddle-point approximation are not
satisfied, that is, we have $\gamma >1$ or we have $x_{\gamma}\ll 1$
for $\gamma<1$, then we can try a complementary approximation
to the integral, with effective expansion parameter $(x_{\gamma})^{1-
\gamma}$. That is, we have an expansion which is good if $x_{\gamma}
\ll 1$ for $\gamma <1$, and $x_{
\gamma} \gg 1$ if $\gamma>1$. In all cases of interest except $\gamma=-1$,
the partition function has a branch-point at the Hagedorn singularity. By
evaluating the discontinuity of the integrand across the cut, we can
transform the $z$-integral (\ref{zetav}) into (for $\gamma$ not an integer)   
\be
\label{cutin}
{1\over x_{\gamma}} \,\int_{0}^{u_1} {\dd u\over \pi} \,{\rm exp}\left\{
-u+ C_{\gamma}\,(x_{\gamma})^{1-
\gamma}\,{\rm cos}\,(\pi\gamma)\,u^{\gamma}\right\} \;\;
{\rm sin }\,\left\{ 
C_{\gamma}\,{\rm sin}\,(-\pi\gamma)\,
(x_{\gamma})^{1-\gamma}\,u^{\gamma} \,\right\} 
,\ee
with
\be
u_1 \equiv (\beta_c -\beta_1)(E-\rho_c\,\Vp) \gg 1,
\ee
within the Hagedorn-singularity-dominance regime. On approximating (\ref{zetav})
by (\ref{cutin}) we are neglecting the contribution of a small circle in
the contour around $\beta=\beta_c$.  This is justified for $\gamma >0$, where
the partition function is bounded near the singularity. The cases $\gamma=-1,0$
will be dealt with exactly below, while $x_{-1/2} \gg 1$ in all practical cases
treated below, so that $\gamma=-1/2$ is calculable in the
saddle-point approximation.    
            
  The integral (\ref{cutin})
admits
a perturbative expansion in $(x_{\gamma})^{1-\gamma}$
with a leading term of
order $1/(x_{\gamma})^{\gamma}$. The result is equivalent to the 
single-long-string
 picture in table (1), with a density of states:
\be
\label{long}
\Omega_{\rm long} \approx f \cdot
{ 
e^{\beta_c \,E +a_c \,\Vp} \over (E-
\rho_c \,\Vp)^{1+\gamma}}
\;\left[1+\ord \left({1\over \Vp (\rho-\rho_c)^2}\right) + \ord \left({f\over
(E-\rho_c \,\Vp)^{\gamma}}\right)\right] 
. \ee
If the critical exponent is an integer, $\gamma=k\geq 2$,  the integral
(\ref{cutin}) is replaced by
\be
\label{cutinn}
{1\over x_k}  \,\int_{0}^{u_1} {\dd u\over \pi} \,{\rm
exp}\left\{
-u+ (x_k)^{1-
k}\,u^k \,{\rm log}\left({E-\rho_c \,\Vp \over u}\right)
\right\} \;\;
{\rm sin }\,\left\{
\pi\,(x_k)^{1-k} \,u^k  
\,\right\}
,\ee
with the same leading behaviour as (\ref{long}), but now with
corrections   
of 
$$
\ord \left((x_k)^{1-k}
\,{\rm log}\,(E-\rho_c \,\Vp)\right)
\sim \ord \left({f\over (E-\rho_c\,\Vp)^k} \;{\rm log}\,(E-\rho_c\,\Vp)\right)
.$$

This regime is equivalent to 
Eq. (\ref{carl}), or table (1) with $\gamma>1$.
 The formally defined temperature $\beta =\partial{\rm
log}(\Omega)/\partial E$ is larger than the Hagedorn temperature.
Although the thermodynamics of these systems is ill-defined, there is
nothing wrong in principle with the density of states (\ref{long}) at
finite volume, and
we will use the `thermodynamic' language and denote these systems
as `non-limiting', or ${\bf NL}[\gamma]$.   

As well as these approximations, there are a number of special
cases which can be evaluated for all $x_\gamma$. The first and
most interesting for our discussion is $\gamma=-1$. Since this
case corresponds to open strings in a finite volume, which is
naturally of central importance to our discussion, we shall
evaluate this case separately in subsection 3.3. The 
two other interesting cases we shall consider now; they are
$\gamma=0$ and $\gamma=1$.

\vspace{0.5cm}
\noindent{\it \underline{Special Case $\gamma=0$}}
\vspace{0.5cm}

The special case $\gamma=0$
admits exact evaluation of the integral (\ref{zetav}), resulting in
\be
\label{gamacex}
\Omega (\gamma=0)_c \sim {1\over \Gamma (f)}\,{e^{\beta_c \,E
+a_c\,\Vp} \over (E-\rho_c\,\Vp)^{1-f}
}\,\left[1 + \ord \left({f\over
\Vt\,(\rho-\rho_c)^2}\right) -\ord \left(e^{-u_1}\right)\right]
,\ee
the first correction factor coming from the analytic terms around the
Hagedorn singularity, and the second one coming from the  next-to-leading
singularity $\beta_1$. 
 This formula interpolates
between the saddle-point result (\ref{sacero}), ${\bf
L}[0]$,
valid at $x_0 = f\gg 1$, and the form (\ref{long}) of ${\bf
NL}[0]$ type, for $f\ll 1$. 

In the marginal case $f=1$,
corresponding to closed strings, 
the limiting or non-limiting behaviour is controlled by the
sign of the correction terms. 
In this case, the singularities are isolated poles of the partition
function, 
 and the sign of the next-to-leading singularity
correction should be negative, leading to a  weakly limiting
behaviour that will be studied in more detail in the next subsection.

\vspace{0.5cm}
\noindent{\it \underline{Special Case $\gamma=1$}}
\vspace{0.5cm}

Finally, we consider the special case $\gamma=1$, where the
parametrization (\ref{zetav}) fails.
 The   singular part of the free energy takes the form
\be
{\rm log}\,Z_{\rm sing} \sim f\,(\beta-\beta_c)\,{\rm
log}\,(\beta-\beta_c)
.\ee
In the saddle-point approximation, the critical point is located
at
\be
{\rm log}\,(\beta_s -\beta_c) = -{E-\rho_c\,\Vp \over f} -1
,\ee
 and the resulting canonical determination of the density of states
is:
\be
\label{cnuno}
\Omega(\gamma=1)_{\rm saddle} \approx {1\over \sqrt{f}} \;
{\rm exp}\,\left\{\beta_c \,E+a_c \,\Vp -f\,e^{-{E-\rho_c \,\Vp  
\over f} -1}-\half \left(1+{E-\rho_c\,\Vp \over f}\right) +\Delta_s
\right\}
.\ee
On evaluating $\beta$, one finds the
marginal case of logarithmically limiting behaviour
seen in table (1). We shall refer to  this
type of density of states as ${\bf L}[1]$.

In fact, the saddle-point approximation has a  limited range
of applicability, since `higher-loop' corrections are of 
$\ord (1/f(\beta_s -\beta_c))$ and thus we may define the saddle-point control
parameter $x_1 \sim f(\beta_s-\beta_c)$: 
\be
\label{gunocond}
x_1 \equiv f\;e^{-{E-\rho_c \,\Vp \over f}} = f\,e^{-\Vt (\rho-\rho_c)}
,\ee
the saddle-point approximation being good for  $x_1 \gg 1$. The analytic
corrections are of order  
$$
{\Delta_s \over x_1} \sim \ord \left(\Vt \,e^{-\Vt (\rho-\rho_c)}\right)
.$$
In the opposite limit, $x_1 \ll 1$, 
 a similar treatment to the one in Eq.
(\ref{cutinn}) gives
\be   
\label{lgun}
\Omega(\gamma=1)_{\rm long} \approx {f\over  
 (E-\rho_c \,\Vp -f
\,{\rm log}\,f\,)^2} \;{\rm exp}\left\{\beta_c\,E+a_c\,\Vp\right\} 
,\ee
which is of `non-limiting' type: ${\bf NL}[1]$, with corrections of
$$
\ord \left({1\over \Vp (\rho-\rho_c -
\Vt^{-1} \,{\rm log} \,(f))^2}\right)+ \ord \left[\left({f\over E-\rho_c\Vp -f{\rm
log}(f)}\right){\rm log}\left({E-\rho_c\Vp -f{\rm log}(f)
\over f}\right)\right]  
.$$

\subsection{Closed Strings in a Finite Box} 

Using these methods, we can now review the results of Ref. \cite{deo}
for closed strings. Let us consider a box of radius $R\geq 1$, with
volume $V_{D-1} \sim R^{D-1}$.
The corresponding analytic structure is given in Eq. (\ref{ploes}),
and we can explicitly evaluate the integral (\ref{clsing}) with the
result  
\be 
\label{expli}
\Omega_{\rm closed} =
\sum_{\alpha} (\beta_{\alpha})^{k_{\alpha}}
\,{e^{\beta_{\alpha}\,E+a_{\alpha}\,V_{D-1}}
\over (k_{\alpha}-1)!}\;
(E-\rho_{\alpha}\,V_{D-1})^{k_{\alpha}-1} 
\;\left[1+\ord \left({(k_{\alpha} -1)(k_{\alpha}-2) \over V_{D-1}
(\rho-\rho_{\alpha})^2}\right)\right]
.\ee 
The corrections come from the analytic terms around each pole, and {\it 
are
absent} for the leading Hagedorn singularity, since $k_c =1$.    
 From the analysis of Ref. \cite{deo}
we learn that the next-to-leading singularity of the partition function
 $\beta_1$ is a pole  
of order $k_1=2D-2$, located at
\be
\beta_c-\beta_1 = \eta/R^2
,\ee
with $\eta\sim +\ord (1)$ in string units.  
The Hagedorn-dominance-rule (H.d.r.) of Eq. (\ref{Hdr}) is given by
\be
\label{otrad}
(\rho-\rho_c')\,R^{D-3} \gg 1
,\ee
which is satisfied for moderately high energy density $\rho>\rho_c' $ and
large radius, provided $D>3$, or for small radius but very high energy
 density $\rho\gg 1$.  In these conditions, the density of states is
approximated by
\be
\label{free4}
\Omega_{\rm closed}    \approx
\beta_c \;e^{\beta_c \,E + a_c\, V_{D-1}}
\left[ 
1-\frac{(\beta_c \,V_{D-1})^{2D-3}}{(2D-3)!}
\,\rho^{2D-3}\;{\rm exp}\left\{-\eta\,R^{D-3}\,(\rho-\rho'_c) \right\}
\right] ,
\ee

Now, in accord with our heuristic argument, in the limit of
large $R$ and provided that the energy density
is larger than the Hagedorn density, the energy dependence of
the microcanonical density of states
{\em always} resembles that for a small compact system
no matter how large the volume we consider. Eq. (\ref{otrad})
is precisely the condition $E\gg \varepsilon_0$,
and tells us the energy scale above
which some of the strings are able to feel the compactness
of the large dimensions by winding.

On using the microcanonical density of states to find the temperature
(from $\beta = \partial \log \Omega /\partial E $) we find
\be
\label{multi}
E \approx \rho'_c V_{D-1} - R^2 \,\log
\left( (\beta_c E)^{3-2 D}\,R^2
(\beta-\beta_c ) \right).
\ee
In particular the specific heat,
\be
C_V \approx
 {R^2\over \beta-\beta_c },
\ee
is always positive and the Hagedorn temperature is approached
monotonically
from below. Therefore, for  closed strings in finite volume,
the Hagedorn temperature is  logarithmically  limiting,
even in the thermodynamic limit for $D>3$. Notice that this limiting
behaviour is very weak, since we can rise the temperature arbitrarily
close to the Hagedorn temperature, while maintaining a moderately low
energy density, provided $R$ is large enough. For future reference,
  we shall denote this 
behaviour as `marginally limiting': ${\bf ML}$.

 When $D\leq 3$  the H.d.r. is violated at sufficiently large $R$.
 In
this case we must use the approximation of non-compact $D_{\infty}-1
 \leq 2$ dimensions,
the saddle-point approximation is good, and we get standard
 ${\bf L}[(D_{\infty}-1)/2]$ behaviour, in agreement with the
canonical ensemble, and the results of table (1). 
 For these lower-dimensional systems there is not enough
energy available to excite winding modes.
   
\subsection{Open Strings on D-branes in a Finite Box}
 
Now we are ready to present new results for D-brane systems. 
Since the critical exponent $\gamma=-1$ and the volume factors
are independent of the ND$+$DN moduli, the discussion can be done in
general for all values of $\nu$, as a function of the number of
large DD directions, $d_{\perp}$, and the volumes $\Vt,\Vp$.  

The control parameter for the saddle-point approximation is  
\be
x_{-1} \equiv x = \Vp\sqrt{\rho-\rho_c \over \Vt}
\ee
and the   H.d.r.  (\ref{Hdr}) for $x\gg 1$ is 
\be
\label{ux}
1\ll {u_1 \over x} \sim \sqrt{\rho-\rho_c}\;\left(\Rt\right)^{-2+d_{\perp}/2}. 
\ee
So, for the very  high-energy regime 
\be
{\rm max}\,\left(\rho_c\,, {\Vt\over\Vp^2}\,, \Rt^{4-d_{\perp}}\right) \ll \rho
\ee
we find ${\bf L}[-1]$ behaviour. In particular, this includes all
systems with $d_{\perp}=0$. 

On the other hand, for $x\ll 1$, the $\gamma=-1$ H.d.r. condition reads
\be 
\label{u} 
1\ll u_1 \sim {(\rho-\rho_c)\,\Vp \over \Rt^2},
\ee
so that the intermediate range
\be
\label{wlreg}
{\rm max}\,\left(\rho_c\,, {\Rt^2 \over \Vp} \right) \ll \rho\ll {\Vt \over
\Vp^2}
\ee
shows a new behaviour characterized by $\gamma=-1$ but $x\ll 1$. In
fact, we can study the density of states at all values of $x$, since
the integral (\ref{zetav}) can be evaluated in closed form  by deforming
the contour to the steepest descent contour as shown in figure 1: 
\be 
\label{omega00}
\Omega(\gamma=-1)_c = \beta_c \,f 
 \;x^{-1} \; I_1(2 x )
\;e^{\beta_c \,E +a_c \,\Vp}\,\left[1+\ord \left({x^2 \over \Vp
(\rho-\rho_c)^2}\right) +\ord \left(e^{-u_1}\right)\right],
\ee
where $I_1$ is the modified Bessel function of the first kind.  
The saddle-point regime $x\gg 1$ is equivalent to (\ref{sa}),
and is therefore of ${\bf L}[-1]$ type, whereas the 
$x\ll 1$ region is marginal from the point of view of the
long string picture (\ref{long}). 
In fact, the exact expression
(\ref{omega00}) leads to rather standard thermodynamics in both extreme
regimes in terms of the parameter $x$, with $x\sim 1$ marking a cross-over
from long-string dominance on the brane to many windings in the transverse
directions.

We can see this when we calculate the temperature. Defining 
\be 
y=\frac{\beta-\beta_c}{f\beta_c},
\ee
we find 
\be 
y=\frac{1}{2x}\left( \frac{I_0(2x)+I_2(2x)}{I_1(2x) } -\frac{1}{x}
\right).
\ee
This function, shown in figure (2), has a tail towards large $x$ giving 
\be 
\beta_c \,E \approx {f\beta_c^2 \over (\beta-\beta_c)^2} .
\ee
In the notation of table (1),  
 this is equivalent to $\gamma=-1$ or $D_o=d_{NN}$.
 It is the behaviour of an open-string system which has all DD
directions small and compact $(d_{\perp} =0)$, 
 with both the energy and specific heat 
diverging as we approach the Hagedorn temperature. 
This was to be expected from the closed-string case, however small 
$x$ gives us a region of different behaviour.

When $x \leqsim 1 $, we find 
\be 
\label{smallx}
E \approx \rho_{c}\,\Vp + \frac{6}{f \beta_c}-12\frac{ (\beta-\beta_c)}
{f^2\beta_c^2}.
\ee
This has a small specific heat as we approach the Hagedorn temperature
provided we can maintain small $x$.
This new type of behaviour is equivalent to keeping the  $\ord (x^2) + \ord (x^4)$
corrections in (\ref{long}), and can be described as `weakly limiting',
due to its small specific heat at Hagedorn temperatures. We shall denote
it by ${\bf WL}[-1]$.

We can obtain a microscopic understanding of this behaviour 
by examining the fraction of strings
which have energy greater than the single-string
 threshold energy $\varepsilon_0 $,  
 and hence are able to wind. 
The energy distribution (\ie the average number of strings in the gas with 
an energy in the interval $\varepsilon \rightarrow \varepsilon 
+ \dd\varepsilon $ 
when the total energy of the system is $E$)
is given by~\cite{deo}
\be 
{\cal D} (\varepsilon;E)\; \dd\varepsilon = 
\frac{\omega (\varepsilon )\Omega(E-\varepsilon)}
{\Omega(E ) }\dd\varepsilon .
\ee
The exponent appearing in $\omega $ depends on the effective dimension 
and hence on whether $\varepsilon $ is greater than or less than  
$ \varepsilon _0 $.
Below $\varepsilon _0$, the distribution is peaked with a power law decay
that is familiar from the closed-string case. 
Above $\varepsilon_0$ and for $x\gg 1 $ we can approximate 
\be 
{\cal D} (\varepsilon;E) = f \;\beta_c \;e^{-\varepsilon\; x/E} .
\ee
Integrating this expression from $\varepsilon_0 $ to $ E $ gives the 
total number of energetic strings;
\be 
x\; e ^{- \varepsilon_0\; x /E}.
\ee
Eq. (\ref{Hdr}) is equivalent to 
$ E \gg \varepsilon _0$ and $x \ll E/\varepsilon $. Thus the total energy must 
always be larger than the nominal threshold energy required to excite 
winding modes.
However the number of energetic strings is roughly proportional to $x$
and for $x \ll 1 $ there are no winding modes, with
all the energy being concentrated in short string excitations
close to the D-brane. As we increase the value of $x$, by raising the 
energy for example, we find that a few of the open strings on the brane
accumulate most of the energy in a manner which is similar to 
closed-string behaviour. As we increase $x$ still further, 
these strings are able to wind, and the spectrum becomes increasingly 
resolved into a low-energy peak and highly energetic winding modes.
The average energy carried by strings 
in the low-energy peak becomes saturated as $x$ becomes large and
$x \approx 1 $ marks a cross-over in behaviour. 

It is interesting to
see what happens if we
 increase the volume of the brane. 
Since $\Omega $ is a monotonically increasing function of $\Vp$, 
there is always an entropic advantage in maximizing it.
Hence the string gas is expected to spread itself over 
the whole available D-brane volume and the volume is expected to 
increase. 
As one might have expected (since there are no winding 
modes in the NN directions), for 9-branes the
result is in accord with Ref. \cite{dienes} 
(based on minimizing the free energy for 9-branes).
However the windings in the DD-directions play an interesting 
role as can be seen by the fact that 
there is an entropic {\em disadvantage} 
in increasing $\Vt$. Winding modes (and more generally 
finite size effects) are seen 
to {\em prevent} DD-directions expanding. A natural
proposal is then that an interplay of these effects 
can stabilize some directions whilst allowing others to expand 
without limit. This is a topic we shall leave for future 
study.

If the H.d.r. with $\gamma=-1$ is not satisfied, \ie if the inequalities
(\ref{ux}) and (\ref{u}) are violated,
 then we must approximate the density of states
by that of a $\gamma_{\infty} = -1 +d_{\infty}/2$ system. In this case,
the control parameter for ${\bf L}/{\bf NL}$ behaviour is
\be
x_{\infty} \sim
\left\{
\begin{array}{ll}
 \Vp\, (\rho-\rho_c)^{\gamma_{\infty} / \gamma_{\infty}-1}
    \;\;\;\;\;\;\;&\gamma_{\infty} \neq 1  \nn\\
\,\nn\\
\Vp \,e^{-(\rho-\rho_c)}        \;\;\;\;\;\;\;& \gamma_{\infty}=1. 
\end{array}
\right.
\ee
One has ${\bf NL}[\gamma_{\infty} \leq 1]$
if $x_{\infty} \ll 1$, and ${\bf L}[\gamma_{\infty}\leq 1]$ if
$x_{\infty} \gg 1$. On the other hand, for $\gamma_{\infty} >1$,
one always has $x_{\infty} \gg 1$ and ${\bf NL}[\gamma_{\infty} >1]$.

There are two main regimes of this kind. The range
\be
{\rm max}\,\left(\rho_c\,, {\Vt\over \Vp^2}\right) \ll \rho\ll \Rt^{4-d_{\perp}}
,\ee
which is only possible for $0<d_{\perp}<4$, leads to ${\bf L}[-1+d_{\perp}/2]$
with the exception of the regime $\rho \gg \Vp$ in $d_{\perp}=3$, which
is ${\bf NL}[1/2]$ instead. Finally, for `low' densities 
\be
\rho_c \ll \rho\ll {\rm min}\,\left({\Vt\over \Vp^2}\,, {\Rt^2 \over \Vp}\right)
\ee
one gets ${\bf NL}[-1+d_{\perp}/2]$ for $d_{\perp}>4$ and ${\bf
L}[-1+d_{\perp}/2]$ for $0<d_{\perp}\leq 4$, with two exceptions: the
regimes $\rho\gg \Vp$ in $d_{\perp}=3$ and $\rho\gg{\rm log}\,(\Vp)$ in
$d_{\perp}=4$, where one finds also ${\bf NL}[-1+d_{\perp}/2]$ behaviour.

\subsection{Summary}

We may summarize the broad lines of our analysis by distinguishing
the two main high-energy limits of interest.  

\vspace{0.5cm}
\noindent{\underline{{\it Extreme High-Energy Limit}}}  
\vspace{0.5cm}

The extreme high-energy regime can be analyzed in simple
terms for all systems. Physically, we expect that at energies much
larger than any scale formed from the T-moduli parameters, the physics
should be similar to the one of a nine-torus of string-scale size and
$\beta_c \,E\gg 1$. 
 Then, the H.d.r. condition (\ref{Hdr})
is trivially
satisfied, since $\beta_1 -\beta_c \sim \ord (1)$. In such a situation, there
is
no clear distinction between NN and DD directions, as both are of the
order of the string scale and are exchanged by T-duality. 
We have   $\gamma=-1$
for the three basic systems of supersymmetric intersections:
$\nu=0,4,8$. Thus, the high-energy regime $\beta_c \,E \gg 1$ corresponds
universally to the ${\bf L}[-1]$ behaviour.  
  Closed strings are described by the ${\bf ML}$ form (\ref{free4}).
 Therefore the entropies
$S={\rm log}(\Omega_c)$  read
\ba
\label{bvcom}
S_{\rm open}  &\approx& \beta_c \,E + 2\sqrt{f\,E} +{\rm const.} -
\ord ({\rm log}\,E) \nn\\
S_{\rm closed} &\approx& \beta_c \,E +{\rm const.}- E^{16} \,e^{-\eta E}
+\ord (E^{15}\,e^{-
\eta E}),  
\ea
with $f, \eta \sim +\ord (1)$.  So, the open-string
 systems clearly dominate the
finite-volume asymptotic entropy:  
\be
\label{hieru}
S_{\rm open} \gg S_{\rm closed} 
.\ee

\vspace{0.5cm}
\noindent{\underline{{\it Thermodynamic Limit}}} 
\vspace{0.5cm}

In the thermodynamic limit, we scale $\Vp \rightarrow \infty$ 
with $\rho\equiv E/\Vp $ fixed and large
in string units. Although this limit does not exist in a strict sense,
 we shall see in section 5 that very stringent conditions can be imposed
on the string coupling constant such that regimes of approximate thermodynamic
behaviour can be defined. So we proceed with the analysis and
 find a large variety of behaviours, depending on
the role played by the winding modes. Roughly speaking, if the winding
modes are sufficiently quenched, one gets agreement with the results
presented in table (1). On the other hand, if the scaling of the DD
directions is such that winding modes store a sizeable portion of the
energy, the behaviour differs from table (1) and one finds a general
tendency towards restoration of limiting behaviour. 

The two main types of behaviour are the single-long-string or ${\bf NL}$
behaviour, with 
 entropy density $\sigma \equiv S/\Vp$ of
 the form (we keep only the leading terms in the thermodynamic limit)
\be
\label{nlen}
\sigma_{{\bf NL}[\gamma]} \approx \beta_c\,\rho-{1+\gamma \over \Vp}\,{\rm
log}\,(\rho)
,\ee
and the various types of `limiting' behaviour ${\bf L}[\gamma]$
 (dominated by a saddle
point) 
\be
\label{len}
\sigma_{{\bf L}[\gamma]} \approx   
\left\{
\begin{array}{ll}
\beta_c\,\rho + 2\sqrt{\rho/ \Vt} 
\;\;\;\;\;\;&{\rm if}\;\;\;\;{\gamma=-1}
  \nn\\
\;\nn\\
\beta_c\,\rho \,+\,{\gamma-1\over \gamma}\,\rho^{\gamma\over \gamma-1}
\;\;\;\;\;\;&{\rm if}\;\;\;\;{\gamma =-\half,\half}
\nn\\
\,\nn \\
\beta_c\,\rho +  {\rm log}\,(\rho)   
\;\;\;\;\;\;&{\rm if}\;\;\;\;{\gamma=0}
\nn\\ 
\,\nn\\
\beta_c\,\rho -e^{-\rho}  
\;\;\;\;\;\;&{\rm if}\;\;\;\;{\gamma=1,}
\end{array}
\right.
\ee
up to positive constants of $\ord (1)$ in string units.

The case $d_{\perp}=0$ is always ${\bf L}[-1]$.
 There is universal agreement with the canonical ensemble results of table (1) for
$0<d_{\perp} <4$, (for example, 
a D$p$-brane in ten dimensions with $p\geq 6$, or a D0--D8 intersection
in
ten non-compact dimensions). For $d_{\perp}> 4$
however (D$p$-branes in ten dimensions  with $p<5$),
 it matters whether the winding modes are quenched, $\Vp \ll \Rt^2$,
 giving
${\bf NL}$ behaviour, or activated into a ${\bf L}[-1]$ system for
$\Vp \geqsim \sqrt{\Vt}$.    
There is an interesting intermediate window, $\Rt^2 \leqsim \Vp \ll
\sqrt{\Vt}$,  of ${\bf WL}[-1]$ behaviour
 (\ref{wlreg}),
with entropy density
\be
\label{wlen}
\sigma_{{\bf WL}[-1]} \approx \beta_c\,\rho + {\beta_c  \,f\over 2} \,\rho
-{\beta_c^2\, \Vp\,f^2 \over 24} \, \rho^2
.\ee
The critical case $d_{\perp}=4$ (a D5-brane in ten dimensions),
is ${\bf L}[1]$, as in table (1)  for sufficiently large transverse
volume, namely $\Vp \ll \sqrt{\Vt}$, and
is ${\bf L}[-1]$ in the opposite regime $\sqrt{\Vt} \leqsim \Vp$.     

Finally, closed strings in the thermodynamic limit have a critical
dimension $D=3$. For $D\leq 3$, we get standard canonical behaviour
as in the table (1), ${\bf L}[(D-1)/2]$. On the other hand, for
$D>3$ winding modes are important enough to turn the naive ${\bf NL}$
behaviour announced in table (1) into the `marginally limiting' 
behaviour ${\bf ML}$ with entropy density
\be
\label{clen}
\sigma_{\bf ML} \approx \beta_c\,\rho -
\rho^{2D-3}\,(V_{D-1})^{2D-4}\,e^{-\rho\,R^{D-3}}
.\ee

We see a tendency of the microcanonical, finite-volume analysis
to restore limiting behaviour in the thermodynamic limit, even if
the naive canonical ensemble determination predicted  non-limiting
features.  
There are a number of exceptions though, in which ${\bf NL}$ behaviour 
survives. Because negative specific heat systems are often signals of
the breakdown of a given microscopic picture, it is worth collecting here
all instances in which they appear. At finite ten-dimensional volume,
   ${\bf NL}$  behaviour
only shows up for moderately `low' energies, such that winding modes
are quenched, \ie the situation is analogous to that of closed strings.
Examples are  transient regimes with $d_{\perp} \geq 3$ (which restricts
to $\nu=0,4$):  
\be
{\rm max}\,\left(\rho_c\,,\,\Vp\,, {\Vt\over \Vp^2}\right) \ll \rho\ll
\Rt
\ee
in a $d_{\perp}=3$ system (a D6-brane). It disappears
if the transverse space is non-compact from the outset. The generic
${\bf NL}$ behaviour appears in the regime 
\be
\rho_c \ll \rho\ll {\rm min}\,\left({\Vt\over \Vp^2}\,, {\Rt^2 \over
\Vp}\right
)
\ee
with $d_{\perp} >4$. The prototype system is a finite D$p$-brane in
non-compact
transverse space with $p<5$ (as in \cite{us}).    
The very high energy-density regime of D6-branes $(\rho \gg 
\Vp)$, and D7-branes $(\rho\gg {\rm log}\,(\Vp))$ in non-compact 
transverse space is also ${\bf NL}$.   
It is important to notice that {\it all} the ${\bf NL}$ regimes 
described here are {\it transients} for finite ten-dimensional volume.
They become truly asymptotic regimes only
 in the case that DD directions are strictly non-compact (which is
unphysical unless we manage to decouple closed strings). 

We would like to note that the ${\bf L}$ behaviour, in which
many long strings form, could indicate that, in an effective
manner, the string tension vanishes.

\section{Thermodynamic Balance}

In this section we consider the behaviour of these systems when two or
 more of them are in thermal contact.
For a given total energy $E_{\rm tot}$, we would like to find the most
probable partition into components $E_i$, with             
\be
\label{encon}
\sum_i E_i = E_{\rm tot}
.\ee
In the free approximation, 
  we must  maximize the total  entropy 
\be
\label{coen}
S(E_{\rm tot})=\sum_i S_i (E_i)
\ee
under the constraint (\ref{encon}). 

We have presented in the
previous sections  approximations
for the functions $S_i (E_i)$, valid in a given   range of energies, so that 
the maximization problem is further restricted by the constraints:
\be
\label{excon}
E_{{\rm min},i} \ll E_i \ll E_{{\rm max},i}
.\ee
Conditions (\ref{encon}) and (\ref{excon})  determine
 the set  $\CD$ where a given form of the entropy functions is valid, and the
maximization problem is defined.
For example, we always restrict our treatment at least to the Hagedorn
regime, $\rho_i > \rho_c$ for all components.  

If a local  maximum exists in the interior  of $\CD$, we have $\nabla S=0$
and the standard equilibrium condition of equality of temperatures,
\be
T_i = T_j
,\ee
with $T^{-1}_i = 
\beta_i = \partial S_i /\partial E_i$.   In addition $\nabla^2 S
<0$,
\ie    the specific heats are positive. If no local maximum is found in the
interior of $\CD$, the total entropy is maximized then on the boundary
$\partial \CD$. In this case, one or several components are pushed to
extreme values of the energy, and the maximization process must be
continued with the extension of the entropy functions to a different
patch.

In principle, thermodynamic equilibrium of a {\it single} component with
 negative specific heat
 is possible when  
it is  sufficiently large in magnitude, $|C^-_V | > C^+_V$,
compared with that of normal systems \cite{thirring}.
  Examples of this situation are
familiar:
large stars and black holes in equilibrium with  radiation in
a finite volume. 

Therefore, it is tempting to suppose that our ${\bf NL}$ systems,
with negative specific heats, can be in equilibrium  with the
normal systems under some conditions. However, the formally computed
temperatures satisfy  
\be
T_{{\bf L}} < T_c
<T_{\bf NL}
.\ee
 Thus, they cannot be in equilibrium and
the maximum entropy must  occur on the boundary of $\CD$.
 In particular, for the combined system with $S_{\rm tot} = S_{\bf L} + S_{\bf
NL}$:   
\be
\label{monon}
{\partial S_{\rm tot} \over \partial E_{\bf NL}} =\beta_{\bf NL} -\beta_{\bf L} <0,
\ee
and $S_{\rm tot}$ is a monotonically decreasing function 
of $E_{\bf NL}$  throughout the entire Hagedorn regime. Therefore, the entropy
is maximized by depleting ${\bf NL}$ energy in favor of the {\bf L} system.  
 Since we  have at least one universal limiting system in any background: 
 the closed strings, we can say that ${\bf NL}$ systems
will have energy densities of the order of the string scale, which is
the minimum energy density for which formula (\ref{nlen}) holds. 
At this point the system of ${\bf NL}$ strings matches to the gas of
massless states at Hagedorn densities. 

Given that ${\bf NL}$ systems will be suppressed, we now turn to
discuss the equilibrium conditions for the various ${\bf L}$ systems. 
From (\ref{len}) we find the energy densities, as a function of
the temperature
\be
\label{enden}
\rho_{{\bf L}[\gamma]} \approx
\left\{
\begin{array}{ll}
\Vt^{-1}\;(\beta-\beta_c)^{-2} 
\;\;\;\;\;\;&{\rm if}\;\;\;\;{\gamma=-1}
  \nn\\
\,\nn\\
(\beta-\beta_c)^{\gamma-1}
\;\;\;\;\;\;&{\rm if}\;\;\;\;{-1<\gamma<1}
\nn \\
\,\nn\\
-{\rm log}\,(\beta-\beta_c) 
\;\;\;\;\;\;&{\rm if}\;\;\;\;{\gamma=1.}
\end{array}
\right.
\ee
So, close to the Hagedorn temperature, the ${\bf L}[\gamma]$ systems with
smaller $\gamma$ have hierarchically larger energy densities. 

The energy density of the ${\bf WL}[-1]$ system
\be
\label{enwl}
\rho_{{\bf WL}} \approx {\Vt\over\Vp^2} - 2{\Vt^2\over \Vp^3}
 \,(\beta-\beta_c) 
\ee
has a maximum of order $\Vt /\Vp^2 \gg 1$, according to (\ref{wlreg}). 
If this limit is exceeded, the density of states of such  systems
takes the ${\bf L}[-1]$ form. So, ultimately, very
 close to the Hagedorn 
temperature, the energy density in the form of ${\bf WL}$ systems is
negligible compared to that in ${\bf L}$ systems.     

The remaining system with a weak limiting behaviour is the  closed-string
sector with $D>3$, whose energy density satisfies
\be
\label{enml}
\rho_{{\bf ML}} \approx (2D-3) R^{3-D} \,{\rm
log}\,\left(V_{D-1} \,\rho_{\bf ML}\right) -R^{3-D}  \,{\rm
log}\,\left(R^2 (\beta-\beta_c)\right)  
.\ee
This system  is by far the weaker limiting system if $R$ is large
enough, since Hagedorn temperatures can be achieved with $\rho  
\geqsim \rho_c \sim \ord (1)$. The same condition for all  the other
systems requires $\rho_{\bf L} \rightarrow \infty$ or $\rho_{\bf WL}\sim
\Vt/\Vp^2 \gg 1$.
 Therefore, we find the following hierarchy of energy densities
in the thermodynamic limit:
\be
\label{hier}
\ord (1)\sim \rho_{\bf NL} \ll \rho_{\bf ML} \ll \rho_{\bf WL} \ll
\rho_{\bf L}
.\ee
Within ${\bf L}$ systems, the one with smaller $\gamma$ wins. Similarly,
within ${\bf NL}$ systems, the one with larger $\gamma$ is expected to
lose energy faster when put in thermal contact with a ${\bf L}$ system.

In the extremely high-energy regime, both open and closed sectors are
limiting. Using the  
forms in  (\ref{bvcom}), 
 we have again a clear dominance of the 
open-string sector. The energies in equilibrium satisfy
\be
\label{energyeq}
E_{\rm open} \approx {e^{2\,\eta\,E_{\rm cl}} \over (E_{\rm cl})^{34}}
,\ee
so that we get the hierarchy
\be
\ord (1)  \ll E_{\rm closed} \ll E_{\rm open},\ee
for $E > \CO (10^2)$. In the conclusions we shall discuss 
the possible implication for models of early cosmology.

There is another situation not covered by the previous discussion
with some theoretical interest. Suppose we can completely decouple
the closed-string and D-brane sectors at all energy scales
(in particular at Hagedorn energy densities). This can be achieved by
taking $N$ D$p$-branes in the large $N$ limit, with $g_s \,N <1$ and
fixed. The effective open-string coupling  remains finite while the
open-closed and closed-closed coupling vanishes. In this case all
extensive quantities in the open-string sector scale to leading order
as $\ord (N^2)$ in the large $N$ limit. Then, if the transverse space
is effectively non-compact and $\gamma_{\infty} >1$ we have a 
truly dominating ${\bf NL}$ 
system, because now it makes sense to concentrate only on the open-string
sector, as totally decoupled from closed strings.  In this situation
we can consider the balance between two such ${\bf NL}$ systems.
 For example, in D0--D4 intersections with infinite DD
directions,   there is a balance between
  $\nu=0$ strings and $\nu=4$ strings, all of them 
 ${\bf NL}$ in infinite transverse
volume.  
 
The maximization of the entropy of such a pair
\be
S(E) \approx \beta_c\,E -(1+\gamma_1)\,{\rm log}\,E_1 -(1+\gamma_2)\,{\rm 
log}\,E_2 
,\ee
under the condition
$ 
 E_1\,,\,E_2 >\ord (1)$,                
at fixed $E=E_1 +E_2$, is achieved with almost all the energy in the
component with smallest $\gamma$. So, in the previous examples,
$\nu=0$ systems with the largest possible world-volume dimensionality
still store almost all the energy even if they
are ${\bf NL}$.    

\section{Beyond the Ideal Gas Regime}

In the previous sections we have studied aspects of Hagedorn regimes in various
string systems, always in the ideal-gas approximation,
 given by the one-loop string
diagrams. The question
 of the effect of interactions in the string thermodynamic ensemble
is a  notorious one.
 From a fundamental point of view, it is  important   
to know what lies `beyond'
 Hagedorn, under the assumption that there is some analogy with
the QCD 
deconfining transition. Namely, is there a phase transition to a regime where
`string constituents' are liberated?

In the
 context of
 fundamental string theories this question is particularly elusive. For
example, the
  presence of gravity automatically invalidates any naive discussion based on
the canonical
 ensemble, or even the microcanonical ensemble in the thermodynamic (infinite
volume) limit.
Gravitational forces have long range and cannot be screened, so that
extensivity cannot be taken for granted. Also, a finite energy
density causes a back-reaction of the geometry.  
For a given total energy $E$, the largest volume that can be considered
approximately static is the Jeans volume:
\be
V_{\rm Jeans} \sim (G_D \,E)^{D-1 \over D-3}
\ee
in $D-1$ spatial
 dimensions, with effective Newton constant $G_D$. The associated length scale
$(V_{\rm Jeans})^{1/D-1}$
 is the equivalent  Schwarzschild radius  for this energy. So,  the 
idea that a black-hole phase lies `beyond' 
Hagedorn (in the sense of large coupling or large energy) is rather natural.
We shall  pursue here this line of thought, without a precise specification
of what the implications would be for a `constituent picture' of the string.

A  step in this direction is the correspondence principle of Ref.
  \cite{polhor}. 
A  wide variety of black holes in string theory can be adiabatically matched to
various perturbative
 string states by appropriately choosing the string coupling constant.
The matching
 point can be
 locally defined by the condition that the supergravity description
of the black-hole
 horizon breaks down due to large $\alpha'$-corrections. Then, both the
mass and 
the entropy
 of the black hole can be matched at this point within $\ord (1)$ accuracy
of the coefficients.
 Under this correspondence, Schwarzschild black holes match onto highly
excited (long)
 fundamental strings, whereas D-branes match onto qualitatively different
kinds of black
 holes depending on the amount of energy on the D-brane world-volume. For a
system of
 $N$ D$p$-branes
 at the same point in transverse space, the classical black-brane
solution is characterized by two radii: a charge radius
\be
(r_Q)^{7-p} = g_s \, N
\ee
in string units, and the standard 
 horizon radius $r_0$.  In the near-extremal regime, $r_0 \ll r_Q$, 
the black-brane
 state matches onto a thermal state of open-string massless excitations,
\ie the 
Yang--Mills gas
 on the D-brane world-volume. On the other hand, in the opposite
 Schwarzschild
limit $r_0 \gg r_Q$,
  the black brane matches onto a long-open-string state on the D-brane
world-volume.
 The correspondence
 principle works in a variety of cases, and agrees with
exact
 microscopic
 determinations of the black-hole entropy in those cases protected by
supersymmetry. 

These arguments
 are intended to apply to finite-energy states of the string theory
 defined by an   
asymptotic Minkowski vacuum. 
In particular, the perturbative long-string states are assumed to be
mostly 
 single-string states. On the other hand, the single-string-dominance picture of
a stringy
  thermal ensemble suggests that perhaps the correspondence principle could be
applied to the
 full thermal ensemble. Strictly speaking, such a statement cannot be
literally 
true because
 of the ill-defined nature of the thermal ensemble in the presence of gravity.
However,  the
 microcanonical density of states calculated for all
Hagedorn
 ensembles  studied in this paper shows a leading linear behaviour for the 
entropy:
\be
S(E)_{\rm Hag} = \beta_c \, E + {\rm subleading,} 
\ee
just as in the single-string picture. 
Based on this,
 we shall propose that the correspondence principle applies provided the
typical thermal state
 in the Hagedorn regime can be defined as an approximately stable
state. The minimal requirement for this is to work at finite volume, well within the
Jeans bound, 
and to use the microcanonical description. One then finds that at sufficiently
high energy, for a given
 fixed value of the string coupling, the most likely state of the system 
is a black-brane or 
black-hole state. The condition that the energy does not exceed the Jeans
bound is then
 equivalent to the requirement that a black hole of that mass fits inside the
given volume. Thus,
 the Jeans bound, when applied to a finite-volume system, becomes
roughly equivalent
 to the holographic bound (a black hole fills all the available volume, and
the corresponding entropy
 gives the maximal information capacity of this background):
\be
\label{holob}
E<E_{\rm Hol} \sim {V^{D-3\over D-1} \over G_D}
.\ee

The interesting aspect of this mild extension of the correspondence principle
 is that the matching point (the transition from the perturbative
string states to
 the black-states) is {\it different} from the Jeans bound. The correspondence
point is defined by the matching of the entropies:
\be
\label{cocu} 
S(E)_{\rm Hag} \sim S(E)_{\rm black}
.\ee
This condition 
defines a critical energy as a function of the string coupling, different from
the Jeans curve.
Naively, this curve is just the transition line between a single long string
and a single black hole. In the thermal gas however, we have seen many instances
in which the thermal ensemble is not dominated by a single long string but
the energy is distributed in a  gas of long strings. In these cases, it may be  
reasonable to apply
 the correspondence principle to each individual string in the thermal
gas. However, these subtleties are irrelevant to 
 the level of accuracy we can  achieve, because the correspondence matching
itself is only known up to $\ord (1)$ factors in the entropy and the energy.
The distinctions between single-long-strings and multi-long-strings (${\bf NL}
$ versus ${\bf L}$), or single- versus multi-black-holes, only show up in
the subleading terms in the entropy, and are thus 
beyond our matching accuracy.

 For a system of $N$ D$p$-branes with longitudinal volume $V_{
\parallel}$ in a torus of
nine-volume $V =
 V_{\parallel} V_{\perp}$, 
there are a number of `phases' that can be identified
on the basis
 of the correspondence
 principle, applied to the typical states both in the bulk
and in the world-volume.
That is, we label a `phase' by those degrees of freedom with
highest entropy, among those that coexist in the thermal ensemble
for a given value of the moduli and the total energy.   

\subsection{Bulk Phase Diagram}

A supergravity gas in ten dimensions has entropy
\be
S(E)_{\rm sgr} \sim V^{1/10} \, E^{9/10}
\ee
and can be matched to a bulk black hole  with entropy 
\be
S(E)_{\rm bh} \sim E \,(g_s^2 \,E)^{1/7}
.\ee
The coexistence
 line $S_{\rm sgr} \sim S_{\rm bh}$  gives a black hole in equilibrium with
radiation in a finite volume, with energy of order
\be
E({\rm sgr}\leftrightarrow {\rm bh}) \sim {1\over g_s^2} \,
 (g_s^2 \,V)^{7/17}
,\ee
and  microcanonical temperature
\be
\label{tha}
T({\rm sgr}\leftrightarrow {\rm bh}) \sim \left({1\over g_s^2
\,V}\right)^{1/17}
.\ee
Since the black-hole-dominated region has negative specific heat, this
temperature is maximal in the vicinity of the transition.
 This configuration is microcanonically stable in finite
volume, in a range
 of energies between the matching point and the Jeans bound \cite{hawking}.

 The graviton
gas can also  be matched
  to a gas of long closed strings. The coexistence curve $S_{\rm sgr} \sim
S_{\rm Hag}$ at temperatures $T\sim \ord (1)$ in string units,
  is independent
of the string coupling and is given by the Hagedorn energy density:
\be
E({\rm sgr}\leftrightarrow {\rm Hag}) \sim V
.\ee
  This Hagedorn phase can be exited at high energy or large
coupling through the
 correspondence curve $S_{\rm Hag} \sim S_{\rm bh}$ (\ref{cocu}):
\be
E({\rm bh}\leftrightarrow {\rm Hag}) \sim {1\over g_s^2} 
,\ee
into a black-hole dominated phase at lower temperatures. The resulting
 phase diagram for
 the bulk or closed-string sector is depicted in figure (3).

An interesting feature of this diagram 
 is the existence of a triple point at the intersection of
 the phase boundaries of
the massless supergravity gas, Hagedorn,
 and black-hole-dominated regimes.  This point lies at Hagedorn
energy density
 $E_c \sim V$, string scale temperatures $T\sim \ord (1)$,
 and considerably weak coupling $g_s \sim 1/\sqrt{V}$ and, somewhat 
optimistically, we would like to interpret its
existence 
 as evidence for completeness of this phase structure. Namely, we are not
missing any major
 set of degrees of freedom. According to this picture, the Hagedorn phase
goes into a 
 black-hole-dominated phase at large energy or coupling, well within the
Jeans or holographic bound:
\be
\label{holbulk}
E<E_{\rm Hol} \sim  {V^{7/9} \over g_s^2}
,\ee
provided we are at
 weak string coupling $g_s <1$.  We see that the Hagedorn regime has no
thermodynamic limit
 whatsoever. If we scale the total energy $E$ linearly with the volume,
we run into the black-hole phase, which ends when   the horizon crushes
the walls of the box (\ie the black hole fills the box).
 Moreover, if the string coupling is larger that $1/\sqrt{V}$, we miss
the Hagedorn phase 
altogether, as the supergravity gas goes into 
the black-hole-dominated phase directly.
In this case, the system has a sub-stringy maximum temperature
\be\label{tma}
T_{\rm max} \sim T({\rm sgr}\leftrightarrow {\rm bh}) < 1.
\ee

\subsection{World-Volume Phase Diagram}

Similar remarks apply to the
 open-string sector in the vicinity of the D-branes.
Here, the details of the correspondence principle depend on the
excitation energy of the D-brane, \ie in the geometric picture,
we must distinguish between the near-extremal $(r_0 \ll r_Q)$ and
non-extremal or Schwarzschild $(r_0 \gg r_Q)$ regimes.

Before proceeding further, it 
 is important to notice that D$p$-branes with $p>6$
cannot be considered as well-defined asymptotic states in weakly-coupled
string theory. The massless fields specifying the closed-string vacuum,
including the dilaton,
grow with transverse distance to the D-brane. As a consequence,
introducing a $p>6$ D$p$-brane in a given perturbative background
inevitably results in a non-perturbative modification of the vacuum
itself. Thus, consistency with the requirement of weak string coupling
throughout the system means that such branes are never far from
orientifold boundaries, and should be better considered as part of the
specification of the background geometry.
In the following, we shall restrict to $p<7$, unless specified otherwise.

  The matching
of the near-extremal $(r_0 \ll r_Q)$ 
black-brane entropy or Anti- de Sitter
 (AdS) throats\footnote{We denote the throat geometries that control the
entropy by AdS, even if they are not strictly AdS unless $p=3$. One
can find a conformal transformation mapping the throat
geometry to an ${\rm AdS}_{p+2}$ for any $p$, \cite{kostas}.} 
\be
\label{sads}
S(E)_{{\rm AdS}_{p+2}} 
\sim N^{1/2} \, 
(V_{\parallel})^{5-p
 \over 2(7-p)} \, g_s^{p-3 \over 2(7-p)} \, E^{9-p \over 
2(7-p)} 
\ee
to a weakly-coupled Yang--Mills gas on the world-volume:
\be
\label{sym}
S(E)_{{\rm SYM}_{p+1}}
 \sim N^{2\over p+1} \, (V_{\parallel})^{1\over p+1} \, E^{p\over p+1}
,\ee
 is the content of the
 generalized SYM/AdS
 correspondence~\cite{malda}, and was studied in detail in
\cite{maldai,us,mart}.

 There are interesting finite-size effects at low temperatures,
$T\leqsim 1/R_{\parallel}$,  in the form of
large $N$ phase transitions of the gauge theory. 
For $p=3$ and spherical topology of the brane world-volume, the
gravitational counterpart is the Hawking--Page transition \cite{hp,w}
between the AdS black-hole geometry and the AdS vacuum geometry
(intermediate metastable phases can be found \cite{dou}). For
our case ($p<7$ and toroidal topology of the branes) the finite-size
effects setting in at the energy threshold $E\leqsim N^2 /R_{\parallel}$ 
are associated to the transition to
 zero-mode dynamics in the Yang--Mills language and  
  to finite-volume  localization
 \cite{grego} in the black-hole language.
At sufficiently low temperatures one must use a T-dual description of
the throat, resulting in an effective geometry of `smeared' D0-branes.
When these D0-branes localize as in \cite{grego} the description
involves an AdS-type throat with $p=0$, which we denote by ${\rm
AdS}_{2}$. In this case of toroidal topology,
 there is no regime of {\it vacuum} AdS dominance,
provided $N$ is large enough \cite{us,pr}.
 We refer the reader to \cite{us,mart,pr}
for
a detailed discussion of such low-temperature phenomena.

  At temperatures $T>1/R_{\parallel}$ these finite-size effects can be
neglected, and the SYM/AdS transition is 
determined by the matching of (\ref{sads}) and (\ref{sym}).
 The transition temperature,
\be
\label{adst}
T({\rm SYM}_{p+1}\leftrightarrow {\rm AdS}_{p+2})
 \sim (g_s\,N)^{1\over 3-p}
,\ee
is smaller than the Hagedorn temperature as long as stringy energy
densities are not reached in the world-volume.

 At this point, it
should be noted that the interpretation of the AdS throats as SYM
dynamics at large 't Hooft coupling (the standard AdS/SYM
correspondence) is problematic for $p=5,6$. For $p=5$, the AdS regime
has a density of states typical of a string theory, with renormalized
tension $T_{\rm eff} = 1/\alpha' \,g_s\,N$ (see \cite{maldaf}).
 For $p=6$ the
qualitative features of the 
thermodynamics of the near-extremal and Schwarzschild regimes are  
essentially the same, so that the boundary $r_0 \sim r_Q$ does not
mark a significant change in behaviour. The holography
properties  required 
to interpret the AdS physics {\it only} in terms of gauge-theory dynamics  
 seem to break down for these cases \cite{malstro,barrab,kutse,us,polpeet}.
  However, the SYM/AdS
correspondence line in the sense of \cite{polhor} can always be
defined, independently of whether there is a candidate 
 microscopic interpretation 
for the entropy (\ref{sads}) in the AdS regime.                

At stringy energy densities $E\sim N^2 \,\Vp$, the SYM/AdS
correspondence line joins the open-string Hagedorn regime.  
The transition from a Yang--Mills gas on the world-volume
to a Hagedorn regime of  open strings $(S_{\rm SYM} \sim S_{\rm Hag})$
occurs at the energy 
\be
E({\rm SYM}_{p+1}\leftrightarrow {\rm Hag}) \sim N^2 \, V_{\parallel}
.\ee
This line joins the 
SYM/AdS correspondence curve at a {\it triple} point 
(see figure (4)), the other phase boundary being the correspondence
curve between the long open strings in the Hagedorn phase, and the
non-extremal black-brane phase.  
Black D$p$-branes in the  Schwarzschild regime
$(r_0 \gg r_Q)$ have  entropy:
\be
S(E)_{{\rm B}p} \sim E\, 
\left({g_s^2 E \over V_{\parallel}}\right)^{1\over 7-p}
.\ee
and match the world-volume Hagedorn phase along the curve:
\be
\label{tr}
E({\rm Hag}\leftrightarrow {\rm B}p) \sim {V_{\parallel} \over g_s^2}
.\ee

Notice that the boundary line    
separating the near-extremal (AdS) and Schwarzschild (B$p$) regimes
of the black branes, given by
$r_0 \sim r_Q$, or   
\be
\label{satrad}
E({\rm AdS}_{p+2}
\leftrightarrow {\rm B}p) \sim N\;{V_{\parallel} \over g_s}
,\ee
also joins the triple point located at $E\sim N^2 \,\Vp$ and $g_s
\,N\sim 1$.   The temperature along this line is
\be
\label{tb}
T({\rm AdS}_{p+2} \leftrightarrow {\rm B}p) \sim \left({1\over g_s
\,N}\right)^{1\over 7-p}
.\ee
This temperature is locally maximal for small energy variations
if $p<5$.  

All these phases lie well
 within the holographic bound, defined by the condition that the
 horizon of the
black brane saturates the available transverse volume:
\be
\label{holwv}
E<E_{\rm Hol} \sim 
{V_{\parallel} \over g_s^2} \cdot (V_{\perp})^{7-p \over 9-p}
.\ee
Notice that this holographic condition is numerically equivalent to the
bulk holographic bound (\ref{holbulk}) for an isotropic box $V\sim R^9$.

 A major difference from the closed-string sector studied in
the previous subsection is the possibility of defining a world-volume
thermodynamic limit, provided the holographic bound (\ref{holwv}) is
satisfied, \ie  if the world-volume energy density $\rho_{\parallel} =
E/\Vp$ satisfies
\be
1\ll \rho_{\parallel} \ll {1\over g_s^2} 
,\ee
there is a thermodynamic limit $\Vp\rightarrow\infty$ with $\Vt$ fixed
and a Hagedorn regime in the world-volume. On the other hand, if
\be
{1\over g_s^2} \ll \rho_{\parallel} \ll {(R_{\perp})^{7-p} \over
g_s^2}
,\ee
there is a thermodynamic limit with the open-string system described
by an infinite black brane.  As pointed out before, the closed-string sector
does not have such thermodynamic limits at fixed coupling,
 and there is no way we can decouple it
completely unless we also scale the string coupling to zero.
Therefore,
the combined system does not have thermodynamic limits at fixed
string coupling, no matter how small. An effective decoupling
of open- and closed-string sectors can be achieved however in the
large $N$ limit with the effective open-string coupling $g_s \,N$ 
held fixed.

\vspace{0.5cm}
\noindent{\it \underline{A Brane Plasma Phase?}}
\vspace{0.5cm}

Notice
 that the transition between the near-extremal and non-extremal
metrics (\ref{satrad}), if continued 
into the Hagedorn regime, leads to a line where the
energy density on the
 D-brane world-volume is of the order of the intrinsic tension of the
brane:
\be
E \sim N\;{V_{\parallel} \over g_s}
.\ee
The accumulation of energy in the D-brane world-volume at the expense of the
bulk is more efficient for the case of limiting D$p$-branes with
$\gamma<1$.    
For these systems,
 the world-volume  energy  density in long open
 strings  near the Hagedorn temperature  
diverges as in Eq. (\ref{enden}),  $ \rho  \sim   
(\beta-\beta_c)^{\gamma-1}$.
 Comparing with the intrinsic   tension \cite{polch}  
 (in string units $\alpha'=1$)
\be
T_{{\rm D}p}
= {1\over (2\pi)^p \,g_s}
,\ee
 we can  see that at 
\begin{equation}
 \left(\frac{ \beta-\beta_c}{\beta_c}\right)^{\gamma-1} \approx
{1\over g_s} 
\end{equation}
the thermal energy is of the same order of magnitude as the rest mass. 
At this point our treatment of D-branes as semiclassical objects,
quantized in a non-relativistic approximation, may break down. So,
any physical picture of this region is neccesarily very conjectural.

A possible  mechanism setting in at these energy densities 
 is the production
of a plasma of branes and antibranes effectively screening
the R-R charge. In principle this screening could be described
by a Higgs-like phase in the low-energy theory of forms.
 However, since we are at weak coupling, 
creating such a plasma would seem to be an 
energetically very expensive way to disperse the charge. Nevertheless
one can see that it might be possible, at least for D-branes
with divergent free energy near the Hagedorn temperature. The
thermodynamic condition for chemical potentials of D-branes $\mu_+$
and anti D-branes $\mu_{-}$  
in equilibrium $ D + \bar D \leftrightarrow X$, where $X$ are massless
NS-NS and R-R fields,   is determined in a standard way 
\begin{equation} 
\mu_+  + \mu_{-} = 0,  
\end{equation}
because chemical potentials for massless fields  in  equilibrium are
zero. If we have a large number of pairs produced we can assume that
$\mu_+  = \mu_{-} = 0$, so our plasma must have zero chemical
potential. Treating branes as some kind of particles with internal
structure (which is given by an excitation spectrum of open strings)
one can  calculate the chemical potential 
\begin{equation}
\mu \sim {\rm log}\,\left(\frac{V_{\perp}}{N}
\sum_{{\vec p}_{\perp}, n} \,e^{-\beta E({\vec p}_{\perp},
n)}\right) 
,\end{equation}
where we have assumed that branes are described by Boltzman statistics
\cite{landau}. One can show that Fermi or Bose statistics are not
going to qualitatively change the picture in the weak-coupling limit.
Here the 
 sum is over all quantum numbers of a single brane including momentum
${\vec p}_{\perp}$ in DD directions and  open-string quantum numbers $n$.  
  $E({\vec p}_{\perp}, n) \approx M_{{\rm D}p} + 
 {\vec p}_{\perp}^{\,2}/2M_{{\rm
D}p} + \epsilon_n$, where
$M_{{\rm D}p}=T_{{\rm D}p} V_{\parallel}$ is the rest
 mass of the D$p$-brane. If $\mu =
0$ we have
\begin{equation}
\frac{N}{V_{\perp}} \sim e^{-\beta M_{{\rm D}p} } \;\sum_{n} 
 e^{-\beta \,\epsilon_n }
\;\int \dd {\vec p}_{\perp} \; e^{-\beta\, {\vec p}_{\perp}^{\ 2}/2M_{{\rm
D}p}}
,\end{equation}
where  $\sum_{n}  e^{-\beta
\,\epsilon_n } = e^{-\beta F_p}$ is a statistical sum of open strings on
the D$p$-brane we are considering. After integration we have
\begin{equation}
\frac{N}{V_{\perp}} \sim e^{-\beta (M_{{\rm D}p} + F_p) }\;
 \left({M_{{\rm D}p}\over \beta}\right)^{d_{\perp}/2}
.\end{equation}
Because both $M_{{\rm D}p}$ and $F_p$ are proportional to $V_{\parallel}$ 
 they both
survive in the thermodynamic limit of large world-volume
 and may 
be suppressed only when $M_{{\rm D}p} + F_p$ is positive. It is clear that for all
systems with divergent $F_p$ (which is negative !) this is not true near
the Hagedorn temperature and we can have   unsuppressed  production of pairs.
 This is the case of  $ p > 6$ branes. The situation with
other branes depends on a balance between $M_{{\rm D}p}$ and $F_p$ -- in a very
similar way (but not exactly 
the same) as for the energy density. We can speculate
that for some still small couplings there is unsuppressed production
of other branes too.  

The present analysis of pair-production processes used a dilute-gas
picture in the transverse directions. Therefore, in the light of
the comments at the begining of section 5.2,  it 
might require very stringent conditions on the
string coupling in order to  consistently apply to $p>6$ D$p$-branes. 
 
\subsection{Thermodynamic Balance}

With these results at hand we can discuss 
 the general features of the thermodynamic
balance between the bulk and world-volume components in the full
weak-coupling parameter space. 

When the combined 
system enters the Hagedorn phase in equilibrium, at temperatures 
of order  $T\sim \ord (1)$, 
our main result in the previous sections 
\be
S_{\rm Hag\;open} \gg S_{\rm Hag \;closed}
,\ee
implies that most of the energy is stored in long open strings, with
the energy density of closed strings kept close to the critical
density $E_{\rm closed} \geqsim V$ (if the volume is large enough).
 Thus, as long as the temperature is
of the order of the string scale, the closed-string sector has a
maximal energy density of the order of the string scale. 
 This means that the system will
have a tendency to exit the Hagedorn phase with most of the
energy concentrated in the world-volume sector, into the black-brane
phase rather than the bulk black-hole phase.   

Since the black-phases have negative specific heat, the maximal
 temperature
of the combined system is achieved in the  Hagedorn
regimes and is $T_{\rm max} \sim \ord (1)$.  
The condition for the combined system to enter the Hagedorn phase
is that the coupling be sufficiently small:
\be
\label{nohag}
g_s < {\rm min}\,\left({1\over \sqrt{V}}\,,\,{1\over N}\right)
.\ee
If the string coupling violates this bound,
the combined system  fails 
to enter the Hagedorn regime and have a sub-stringy
maximal temperature as we increase the total energy. The maximal
temperature of the bulk is
\be
T_{\rm closed\;max} \sim \left({1\over g_s^2 V}\right)^{1/17}
<1\ee
if (\ref{nohag}) holds, 
while the maximal temperature of the boundary sector is  
\be
T_{\rm open\;max} \sim {\rm
max}\,\left((g_s\,N)^{1/3-p}\,,\,(g_s\,N)^{1/p-7}\right)
<1.\ee
Since both sectors are supposedly in equilibrium, which of the 
two maximal temperatures is attained depends on the
detailed values of the moduli and coupling.    

For phenomenological applications based on weakly-coupled brane-models
we always demand the brane sector to be perturbative, due to the 
phenomenological requirement of weak gauge couplings in the Standard
Model gauge group, and the technical requirement of calculability. So,
in this context we always work with brane sectors satisfying $g_s
\,N<1$. Under these conditions, the bound (\ref{nohag}) depends only
on the total volume for closed-string propagation. The critical point
at $g_s^2 \,V \sim 1$ was derived in subsection 5.1 for a
nine-dimensional box. In fact, it 
 is independent of the number of large dimensions
available for closed-string propagation. The equilibrium between
a supergravity gas in $D$ space-time dimensions with entropy
\be
S(E)_{\rm sgr} \sim (V_{D-1})^{1\over D} \,E^{D-1\over D}  
\ee
and a black hole, 
\be
S(E)_{\rm bh} \sim E \,(g_s^2 E)^{1\over D-3}
,\ee
occurs at energies
\be
E({\rm sgr}\leftrightarrow {\rm bh}) \sim {1\over g_s^2}\,
\left(V_{D-1}\right)^{D-3\over 2D-3} 
,\ee
and the resulting maximal temperature is of order
\be
T({\rm sgr}\leftrightarrow {\rm bh}) \sim \left({1\over g_s^2
\,V_{D-1}}\right)^{1\over 2D-3} 
,\ee
leading to the same critical coupling for all values of $D$.
 Thus, we see that a  
cosmological weakly-coupled 
Hagedorn regime, with temperatures of $\ord (1)$ in string units, 
 is only possible for a sufficiently small
universe. We would be led then to a scenario of the type studied in
\cite{brandvafa}, with the difference that 
 open strings dominate the
Hagedorn regime.   

\section{Concluding remarks}

In this paper we have presented a general study of the thermodynamics
of strings propagating in  D-branes backgrounds.  
Particular attention has been paid to the 
Hagedorn regime and the associated long-string behaviour
at weak string coupling where the free (ideal
gas) approximation is accurate. 

We have stressed that, in any well defined ensemble, winding 
modes are central to the behaviour of the system. This is 
because the classic problems one faces in defining a
satisfactory thermal ensemble (the Jeans instability for example)
can only be bypassed by working at finite volume. 
Only at finite volume does it make sense 
to define an approximate thermodynamic limit.

We discussed the thermodynamic behaviour in a toroidally 
compactified space by applying the powerful calculational
techniques of Ref. \cite{deo} to general string sectors in D-brane
backgrounds. As expected we find that a pivotal role is played by the effect
of winding modes. For example, in a gas of open strings 
on an isolated D-brane, if the volume of large winding-supporting dimensions
(DD directions) is sufficently large compared to the world-volume (NN
directions), namely $\Vp < \sqrt{\Vt}$, then the windings are
`deactivated' and the thermodynamics is well described by the
non-compact approximation \cite{us}, with limiting behaviour for 
D$p$-branes with $p\geq 5$, and non-limiting, negative specific heat
for $p<5$. On the other hand when 
$\Vp > \sqrt{\Vt}$ windings tend to be
`activated' and the Hagedorn temperature is limiting with positive
specific heat. Moreover, the Hagedorn behaviour switches on
when a critical density is reached on the brane (or intersection).

We also compared the thermodynamics of different systems 
and showed that those which are approximated by a non-limiting density of
states are thermodynamically subleading compared to limiting systems.
Since closed strings are a universal limiting system (in a finite volume), 
we find that all non-limiting transient behaviour is suppressed 
in the full thermal ensemble. 
Also we found that, in a given intersection,
 open strings on parallel branes ($\nu=0$) 
of the largest dimensionality 
dominate the thermodynamics.   

Armed with the correspondence principle of Ref. \cite{polhor} and its
generalizations, we ventured into the speculative terrain beyond the
ideal-gas approximation. Under some assumptions, we were able to 
build a consistent qualitative phase diagram, which used the degrees
of freedom present in weakly-coupled string theory (fundamental
strings and D-branes), together with the necessary ingredient to
satisfy the holographic bound (\ie a black-hole-dominated phase).
The  main feature
of these phase diagrams 
is that Hagedorn phases are bounded by black-hole-dominated
phases at large energy or coupling.   

A remaining bone of contention is the fact that the non-relativistic
approximation for the D-brane quantization could break down within the
Hagedorn regime, before the energy density is large enough to match
to a black brane. This question might depend on a detailed study
of the contribution of collective coordinates of the D-branes to the
thermodynamics. Such a study is for the moment beyond our
capabilities, although some ingredients were laid out at the end of
section 5. We argued that, for sufficiently small string coupling,
a plasma of brane-antibrane pairs might be produced, with local
screening of R-R charge. 

Our results  may have some relevance to cosmology
of D-brane backgrounds so let us conclude with some 
more speculative remarks concerning earlier and possible future work
in this area.
One interesting issue which was discussed in the context of 
closed strings was the possibility of explaining 
the choice of four dimensions from Hagedorn behaviour. 
How does the presence of D-branes effect this question?
First, we should emphasize that the dominance of open strings 
is true even in the most extreme case 
of an initially  small universe where all the dimensions are 
of $\CO (1)$. Such a case was examined in Ref. \cite{brandvafa} 
where it was suggested that a small universe dominated by 
closed-string winding modes would have been unable to expand 
unless the winding modes annihilated. For closed strings, 
annihilation is virtually impossible unless $D\leq 4$ space-time 
dimensions are large. Here we have seen that when there are
D-branes, at least in the 
weak coupling approximation that we have been using here, 
most of the energy flows into open strings on the brane even for modest 
energies. In addition, it is only possible to be in the long-string regime 
at very weak coupling where the cosmology is expected to 
be dominated by the tension of the D-brane itself. Hence there seem
to be no regions where this type of scenario is applicable.

This is probably the appropriate point to introduce 
an alternative (and equally speculative) idea 
to explain four (or at least the low number of) space-time dimensions.
This might be called the `melting' scenario. 
We have seen that when the DD directions are large, the 
thermodynamic behaviour is very different for $p< 5$ and 
$p\geq 5$ branes. Hence, in a universe full of different dimensionality 
D-branes, the higher-dimensional limiting 
D-branes attract all the energy of the system until they 
`melt', when the energy density in
their world-volumes is sufficiently large. This would leave only 
the low-dimensional $p<5$ branes 
which, as we have seen, are non-limiting in the Hagedorn regime 
(\ie they can 
be close to the Hagedorn temperature with only a finite energy density)
and therefore able to survive.
There are however two possibly fatal objections to this scenario.
First, for it to make sense one should
be able to build the D-branes themselves as `bound states' of
fundamental strings. The R-R charge could disappear if it has only
a dynamical, low-energy, meaning. A nice example of this would be
the thermal `un-wrapping' of the magnetic charge of a 
't Hooft--Polyakov monopole gas at sufficiently high temperatures. 
Unfortunately, in the
case of R-R charge and D-branes, we have no evidence that the
corresponding conserved charges can be dynamically `unwrapped', at
least in a context where the bare string coupling is kept small. The
reason is that R-R charges are related to Kaluza--Klein momenta through
dualities, and the `unwrapping' of Kaluza--Klein momenta requires some  
non-perturbative background dynamics (topology change). As well as 
this technical problem, there is a more serious conceptual 
problem with the `melting' idea. If we are
willing to believe that D6-branes `melt' as the universe contracts and
energy densities become very large, we have to accept that D6-branes
can `condense' out as the universe expands and cools down from a 
very dense stage. Hence, without a rigorous knowledge of what happens to the 
D-branes after they `melt', this idea remains extremely speculative. 

More generally, however, it is  clear that 
the Hagedorn regime is a rich source of cosmological possibilities
and in particular gives an
interesting new kind of disequilibrium; a D-brane `formed' 
in a hot gas of closed strings will inevitably attract all 
the bulk entropy onto its surface.  Any new source of 
disequilibrium is of great interest for both baryogenesis and inflation.

\subsection*{Acknowledgements}
We would like to thank Karim Benakli, 
Keith Dienes, Emilian Dudas, Sanjay Jain,  
Keith Olive, Graham G. Ross,  
Subir Sarkar and Miguel A. V\'azquez-Mozo for conversations.
The work of E. R. is partially supported by the Israel Academy of
Sciences and Humanities Centres of Excellence Programme, and the
American Israel Bi-National Science Foundation.

\newpage

\begin{figure}
\vspace*{1in}
\hspace*{0.3in}
\epsfysize=4.in
\epsffile{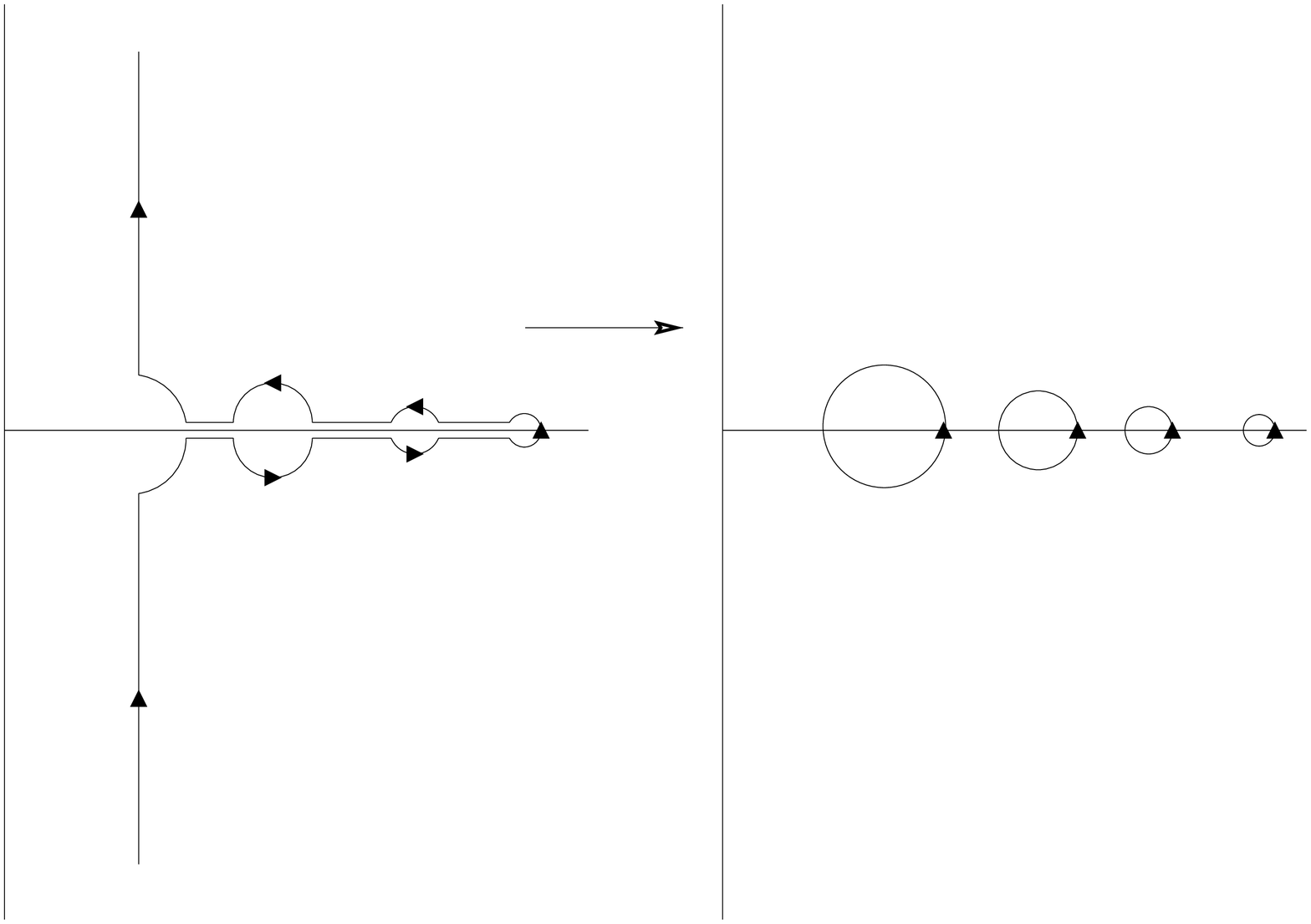}
\caption{Contour deformation for isolated singularities.} 
\end{figure}

\newpage

\begin{figure}
\vspace*{-2in}
\hspace*{-0.5in}
\epsfysize=12in
\epsffile{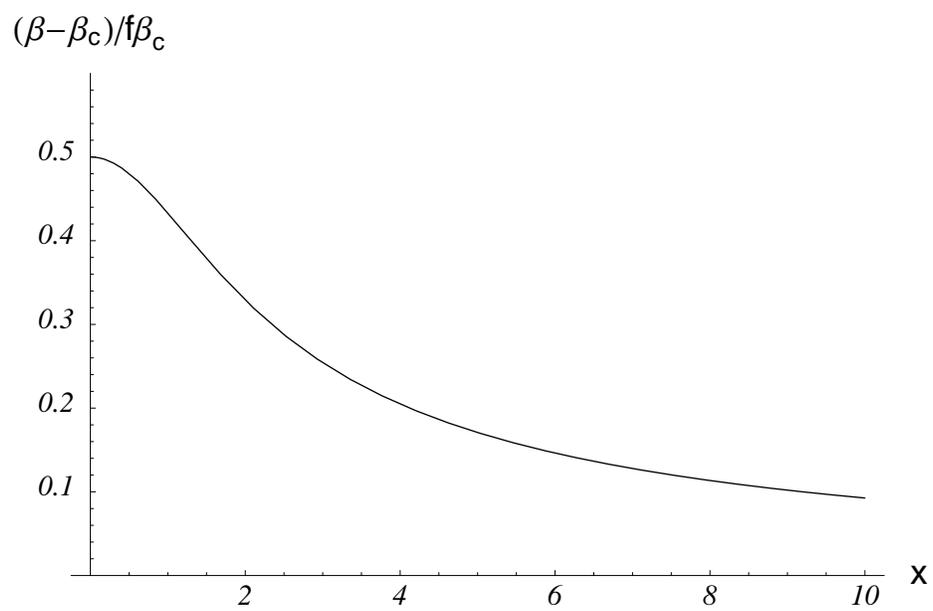}
\vspace{-6in}
\caption{Variation of $\beta-\beta_c$ with $x=\sqrt{f
\, (E-\rho_{c}
\,\Vp)}$ for open strings.} 
\end{figure}

\newpage
\vspace*{1in}
\begin{figure}
\hspace*{0.5in}
\epsfxsize=5in
\epsffile{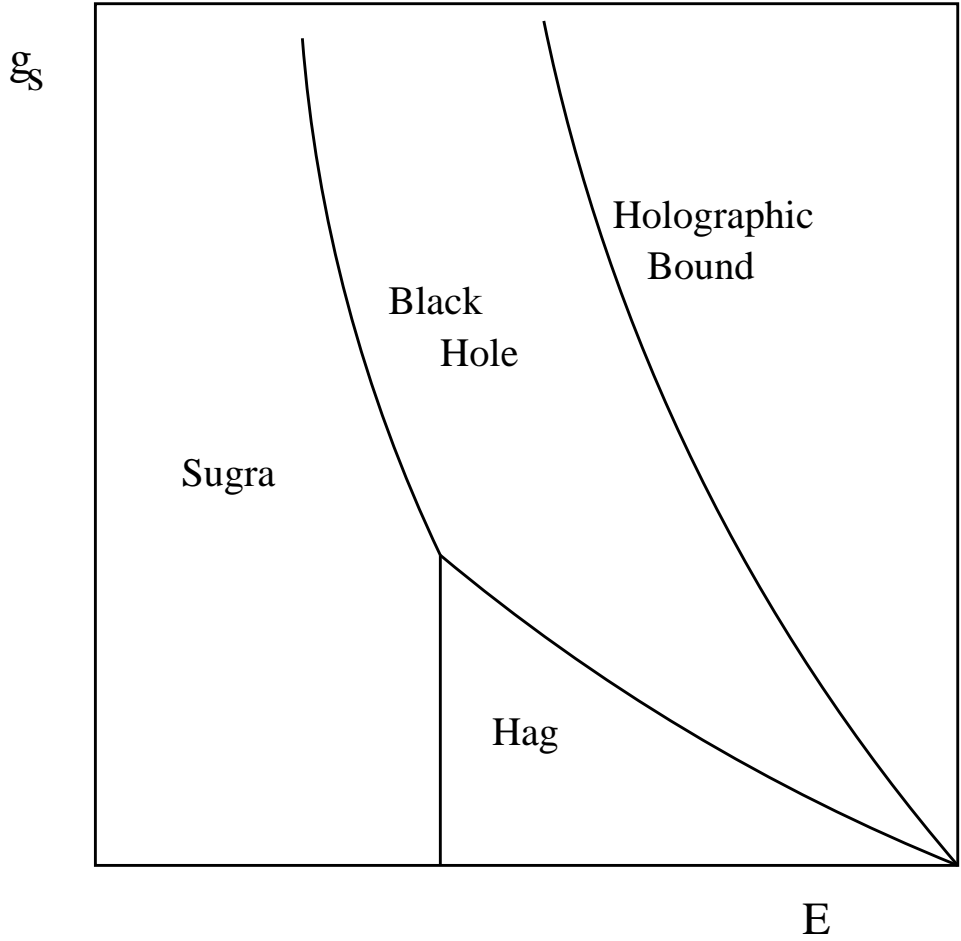}
\caption{
Bulk phase diagram. Only the region $g_s <1$ is represented. The triple
point separating the supergravity gas, black hole, and
 Hagedorn-dominated regimes 
is located at $g_s \sim 1/\sqrt{V}$, and $E\sim V$. The rightmost region
is excluded by the holographic bound (\ref{holbulk}).}
\end{figure}

\newpage
\begin{figure}
\hspace*{0.5in}
\epsfxsize=5in
\epsffile{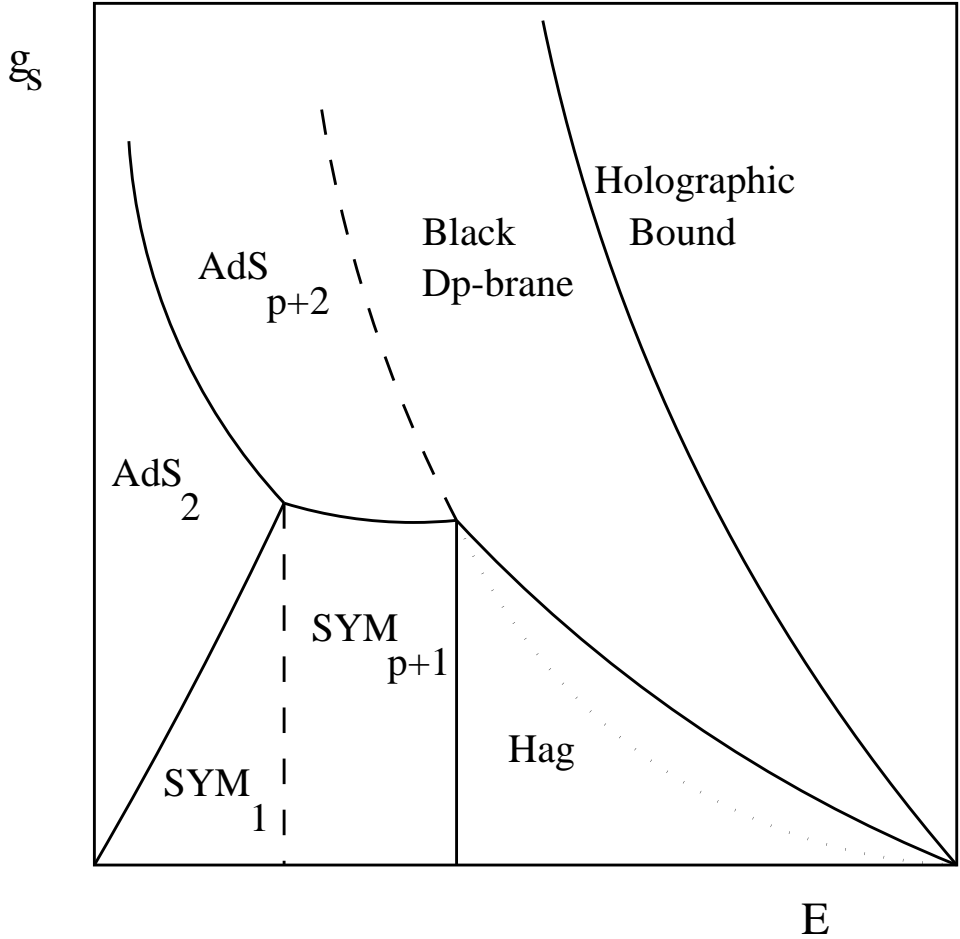}
\caption{World-Volume phase diagram for $g_s <1$.
 Thick lines represent semiclassical phase transitions or
correspondence curves with a major change in the degrees of freedom,
whereas dashed lines represent smooth cross-overs within the same
basic description.  The triple point at low energies was
studied in \cite{us}; it lies at
$g_s \sim R_{\parallel}^{p-3}/ N$, $E\sim N^2 / R_{\parallel}$, and
is due to finite-size effects in the Yang--Mills theory. 
 The triple point
at Hagedorn energies is located at $g_s \sim 1/N$ and $E\sim N^2 \,\Vp$. The 
dotted line within the Hagedorn region 
 represents the D-brane bare-mass threshold,   
$E\sim N\,\Vp/g_s$.     
 Again, the rightmost region is excluded by the holographic
bound (\ref{holwv}).}
\end{figure}
\end{document}